\documentclass{IEEEtran}
\IEEEoverridecommandlockouts
\usepackage{cite}
\usepackage{amsmath,amssymb,amsfonts}
\usepackage{algorithmic}
\usepackage{amssymb}
\usepackage{graphicx}
\usepackage{textcomp}
\usepackage{xcolor}
\usepackage{array}
\usepackage[justification=centering]{caption}
\usepackage[caption=false,font=normalsize,labelfont=sf,textfont=sf]{subfig}
\usepackage{textcomp}
\usepackage{stfloats}
\usepackage{url}
\usepackage{verbatim}
\usepackage{soul}
\usepackage{soul, color, xcolor}
\usepackage{booktabs}
\usepackage[numbers]{natbib}
\usepackage{threeparttable}
\usepackage[none]{hyphenat}
\def\BibTeX{{\rm B\kern-.05em{\sc i\kern-.025em b}\kern-.08em
    T\kern-.1667em\lower.7ex\hbox{E}\kern-.125emX}}
\def\BibTeX{{\rm B\kern-.05em{\sc i\kern-.025em b}\kern-.08em
    T\kern-.1667em\lower.7ex\hbox{E}\kern-.125emX}}
\usepackage{balance}

\begin{document}
\title{
Tensor Train Decomposition Based Noise Reduction and Enhanced Parameter Estimation for FMCW MIMO Radar Systems
}

\author{\IEEEauthorblockN{Luoyan Zhu$^{\flat,\S}$, Sergiy A. Vorobyov$^{\S}$, Jie Wang$^{\flat,\S}$, Yinsheng Liu$^{\flat}$, Zhangdui Zhong$^{\dagger,\sharp}$}
\IEEEauthorblockA{
{$^{\flat}$}{School of Electronic and Information Engineering, Dalian Maritime University, Liaoning, China 116026}\\
{$^{\dagger}$State Key Laboratory of Rail Traffic Control and Safety, Beijing Jiaotong University, Beijing, China 100044} \\
{$^{\sharp}${Collaborative Innovation Center of Railway Traffic Safety}}\\
{$^{\S}$Department of Information and Communication Engineering, Aalto University, Finland}\\
}
}
\author{Luoyan Zhu,~\IEEEmembership{Member,~IEEE,}
        Sergiy A. Vorobyov,~\IEEEmembership{Fellow,~IEEE,}
        Jie Wang,~\IEEEmembership{Senior Member,~IEEE,}
        Yinsheng Liu,\
        and Zhangdui Zhong,~\IEEEmembership{Fellow,~IEEE}

\thanks{
(\emph{Corresponding author: Jie Wang.})
Luoyan Zhu and Jie Wang are with the School of Information Science and Technology, Dalian Maritime University, Dalian 116026, China
 (e-mail: ly.zhu@dlmu.edu.cn;
 wang\_jie@dlmu.edu.cn).

Sergiy A. Voroboyv is with the Department of Information and Communications Engineering, Aalto University, 02150 Espoo, Finland (e-mail: sergiy.vorobyov@aalto.fi).

Yinsheng Liu, and Zhangdui Zhong are with the School of Electronic and Information Engineering,
Beijing Jiaotong University, Beijing, 100044, China (e-mail: ys.liu@bjtu.edu.cn; hedanping@bjtu.edu.cn; kguan@bjtu.edu.cn; zhdzhong@bjtu.edu.cn).

}}
{}
\maketitle

\begin{abstract}
Frequency modulated continuous wave (FMCW) radar is widely used in autonomous driving and industrial inspection due to its high-resolution target location and velocity estimation capability.
 However, the plethora of connected devices in automotive applications introduces electromagnetic interference and brings challenges to location-aware services, primarily due to the issue of low signal-to-noise ratio (SNR) caused by mixed noise contamination.
 Conventional matrix-based signal processing methods exhibit performance deterioration when handling higher-order signals under low SNR conditions.
To address this challenge, this paper proposes a tensor decomposition-based framework that jointly performs noise reduction and parameter estimation for four-dimensional signals in FMCW multiple-input multiple-output (MIMO) radar systems.
Specifically, the framework exploits the inherent low-rank structure and multidimensional correlations of the received signals through tensor train decomposition to effectively separate noise subspace.
 A data smoothing processor then reconstructs an augmented signal tensor to resolve rank deficiency caused by coherent signals.
 Finally, an enhanced rotational subspace algorithm is employed to jointly decouple the distance, velocity, and angle parameters by exploiting the structural fitting to the restored signal.
Both simulation and field experiments under real-world noise demonstrate that our proposed framework achieves significant noise reduction while improving target SNR and parameter estimation accuracy.
These advancements make the proposed framework a robust solution for high-precision MIMO FMCW radar applications in dynamic, noise-polluted environments.
\end{abstract}

\vspace{-1em}
\section{Introduction}
Enabling advanced location-aware services in vehicle-to-everything (V2X) networks requires the real-time sensing capabilities with high accuracy and low power consumption to achieve effective information interaction \cite{he2023physics,10457955,20214D}.
Over the recent years, millimeter-wave (mmWave) frequency-modulated continuous wave (FMCW) signals are prevalent in modern automotive radar systems, offering a cost-effective and easily implementable solution for high-resolution ranging and velocity estimation \cite{10044244, 9760104}.
 The fundamental principle of FMCW radar lies in transmitting a linearly frequency-varying signal over a chirp duration.
 When the transmitted signal reflects from a target, the received echo exhibits phase and frequency shift induced by delay and Doppler effect.
 By analyzing the intermediate frequency derived from mixing the transmitted and received signals, precise target distance and velocity can be extracted.

 Due to its ability to simultaneously resolve multiple targets with high accuracy and low hardware complexity, FMCW radars are particularly suitable for automotive and industrial applications.
However, the low transmitted power of FMCW radars with single antenna at mmWave band results in weak target echoes buried in heavy background noise.
To support high-resolution in dynamic environments, researchers have been working on noise reduction methods for different types of radars.
Hu et. al in \cite{9853513} have introduced a denoising method based on neural networks for pulse radars at a low SNR, leading to the recognition accuracy of radar waveforms being improved.
Hao et. al in \cite{9978673} have proposed an automatic mode selection scheme for ground penetrating radar (GPR) signals denoising with poor SNR, revealing a dipped return signal at great depths in the real GPR data.
Tan et. al in \cite{9521673} have proposed a novel end-to-end self-supervised denoising model for synthetic aperture radar (SAR), enhancing the spatial details significantly.
Conventional signal processing methods for FMCW signals, such as fast Fourier transform (FFT), remain widely adopted for range and velocity estimation owing to their low computational complexity.
 However, FFT-based approaches suffer from compromised resolution due to side-lobe leakage, rendering them ineffective for high-precision applications.
 Furthermore, while FFT excels in two-dimensional (2D) parameter estimation, it fails to support accurate direction-of-arrival (DOA) estimation, a prerequisite for multi-target localization in cluttered environments.
 Such situation poses challenges for further applications, such as target detection \cite{10164024}.

To enhance the parameter estimation and localize targets in angular domains, multiple antenna elements at both transmit and receive sides, namely multiple-input-multiple-output (MIMO), are used to construct a large virtual aperture for a better angular resolution by exploiting spatial degrees of freedom \cite{xu2021transmit,9495264}.
FMCW signals can be combined with MIMO arrays, leading to additional signal dimensions in azimuth and elevation spatial domain.
Such configuration improves the signal-to-noise ratio (SNR) and enhances the capability of parameter estimation.
 Subspace-based methods like multiple signal classifier (MUSIC) \cite{1143830} and the estimation of signal parameters via rotational invariance technique (ESPRIT) \cite{32276} leverage spatial spectrum analysis or rotational invariance principles, which match well with MIMO configuration.

 These techniques enhance resolution beyond the Rayleigh limit but rely on matrix-based signal representations, discarding inherent multidimensional correlations in multidimensional radar signals.
 For instance, MUSIC requires exhaustive 2D spectral searches, while ESPRIT exploits rotational invariance in matrix structures but overlooks higher-order signal dimensions.
 Consequently, these methods struggle to jointly estimate range, velocity, and angle parameters, particularly in multi-target scenarios with dynamic interactions.

To address these challenges, this paper proposes a novel tensor-based signal processing framework that naturally deals with multidimensional signals.
 Firstly, we develop an efficient noise reduction method for four-dimensional (4D) signals to meet the low-complexity and high-adaptability requirements of higher-order signal processing.
Building upon this foundation, we further present an improved joint parameter estimation algorithm that enhances estimation accuracy by leveraging the inherent structure of tensor signals.
Specifically, an FMCW MIMO radar system is considered in this paper for 77~GHz automotive applications, where both the transmitting and receiving sides are equipped with uniform rectangular arrays (URAs) used for estimating parameters over range, velocity, azimuth and elevation angles of targets.
The structural correlations of multidimensional radar signal is modeled by a low-rank canonical polyadic decomposition (CPD), which is particularly suited for Vandermonde-structured signals to capture their global inter-dimensional relationships through factor matrices.
However, conventional CPD implementations using alternating least squares (ALS) face prohibitive computational complexity for high-order tensors and show sensitivity to local minima in noisy scenarios.
Moreover, the global CPD rank remains suboptimal due to the deficiency in capturing the local inter-dimensional relationships, which leads to residual noise entangled in redundant subspaces.
To address this, we propose a tensor train (TT) decomposition framework originated from the functional equivalence, which we also developed here between CPD and TT under Vandermonde constraints.
  It enhances noise suppression by replacing CPD's rigid global rank with TT's adaptive multi-scale rank selection criterion.
We also develop a TT-recompression method for signal in CPD-to-TT format to extract latent lower-rank structure of CPD, which corresponds to the noise removal across different tensor modes and enables comparable denoising performance with TT framework.
To mitigate rank deficiency from coherent signals after noise removal, we design a 4D data smoothing processor for exploiting multi-linear signal structure and extend the conventional matrix-based ESPRIT to higher-order utilization for joint parameter estimation.

Specifically, we first exploit the global correlation among different dimensions of the associated Vandermonde-structured signal tensor  with the CPD model.
By reformulating CPD, TT is derived to exploit CPD's latent lower-rank structure by replacing CPD's rigid global rank with TT's adaptive multi-scale rank constraints.
To automatically determine optimal TT ranks, we propose an adaptive rank selection scheme for noise reduction based on minimum description length (MDL), which avoids the prior knowledge of noise levels, to set the truncation threshold in conventional TT decomposition.

Second,
to address rank deficiency caused by coherent signals, we deduce a pre-processing data smoothing method by employing the forward-backward averaging (FBA) for the 4D subspace algorithm.
For that, we extend rotational invariance principle from conventional single-dimensional ESPRIT to multidimensional signal structures to achieve the robust higher-order parameter estimation.
The generalization enhances the joint parameter estimation accuracy by handling multi-linear signal representations and decoupling the multidimensional relationships.

 Finally, we conduct computer simulations and field measurements to demonstrate the performance of our method from analytical and practical perspectives.
The proposed denoising method based on TT-MDL, as well as the conventional CPD and recompressed CPD, are implemented for 4D FMCW signals.
Our simulations illustrate that the proposed method can remove noise components more efficiently with a lower computational complexity.
The resulting SNR gains further lead to improvement of target detection performance and parameter estimation accuracy.
Moreover, we also conduct field measurements and demonstrate the effectiveness and practicability of our proposed method.

\subsection{Contributions}
The main contributions of this work are as follows:
\begin{itemize}
 \item \textbf{Noise reduction based on TT decomposition}:
 To extract low-rank structures in adaptive multi-scale denoising, we propose a TT decomposition-based approach with integrated MDL criterion for optimal TT rank selection.
  The objective is also to maintain computational efficiency.
  Moreover, we develop a QR-enabled recompression mechanism to validate and optimize the rank truncation process to address the limitations of conventional CPD decomposition.



 \item \textbf{Enhanced parameter estimation}:
 To address the rank deficiency problem resulting from coherent signals, we deduce a pre-processing data smoothing that facilitates the 4D ESPRIT-based parameter estimation, and exploits the rotational invariance property of the signal tensor to decouple the multidimensional relationships and enable accurate 4D target information extraction.

 \item \textbf{Practical and experimental validation}:
 To evaluate the effectiveness of the proposed framework, we conduct practical measurements and extensive simulations under various SNR conditions.
 The results demonstrate the performance improvement and significant practical value of the proposed high-dimensional method.
\end{itemize}

\subsection{Notations}
Vectors and matrices are denoted in the paper by bold letters in the form of $\boldsymbol{x}$ and $\boldsymbol{X}$, respectively.
A tensor of order $N$ is denoted by $\boldsymbol{\mathcal{X}}\in \mathbb{C}^{I_1\times I_2 \times ... \times I_N}$ and its $(i_1, i_2,..., i_N)$-th entry by $x_{i_1, i_2,..., i_N}$.
The order of a tensor is defined as the number of dimension indices or modes.
The symbols $\boldsymbol{1}_n\in \{1\}^{n\times1}$, $\boldsymbol{I}_n\in \{0,\, 1\}^{n\times n}$, and $\boldsymbol{J}_n \in \{0,\, 1\}^{n\times n}$ stand respectively for the all-ones vector, the identity matrix, and the exchange matrix, i.e., the matrix with 1's on the anti-diagonal and 0's elsewhere,
that satisfies $\boldsymbol{J}_n\boldsymbol{J}_n=\boldsymbol{I}_n$.
The identity tensor is expressed as $\boldsymbol{\mathcal{I}}_{N,R}$, where $N$ is the order of the tensor, and $R$ is the number of elements in each dimension.

A submatrix of matrix $\boldsymbol{U}$ formed from its first $n$ rows is denoted as $\boldsymbol{U}^n$.
The mode-$n$ product of the tensor $\boldsymbol{\mathcal{X}}$ and a matrix $\boldsymbol{U} \in \mathbb{C}^{R \times I_n}$ is denoted as $\boldsymbol{\mathcal{X}} \times_n \boldsymbol{U} \in \mathbb{C}^{I_1\times ... \times R\times ... \times I_N}$.
The contraction product between two tensors $\boldsymbol{\mathcal{A}}\in \mathbb{C}^{A_1\times A_2 \times ... \times A_P}$ and $\boldsymbol{\mathcal{B}}\in \mathbb{C}^{B_1\times B_2 \times ... \times B_Q}$, where $A_p=B_q$, $p=1,...P$, $q=1,...,Q$, gives a tensor of order $P+Q-2$, and is denoted as $\boldsymbol{\mathcal{A}}\times^q_p \boldsymbol{\mathcal{B}}$.
Moreover, $\sqcup_p$ denotes the contraction of tensors $\boldsymbol{\mathcal{A}}$ and $\boldsymbol{\mathcal{B}}$ along the $p$-th mode, i.e., $[\boldsymbol{\mathcal{A}}\sqcup_p \boldsymbol{\mathcal{B}} ]$.
The operator $\mathrm{reshape}$ that enables the relationship $\boldsymbol{\mathcal{B}}=\mathrm{reshape}(\boldsymbol{\mathcal{A}},[B_1, B_2,..., B_Q])$ maps the $(a_1,a_2,...,a_P)$-th element
of $\boldsymbol{\mathcal{A}}$ to the $(b_1,b_2,...,b_Q)$-th element
of $\boldsymbol{\mathcal{B}}$, where
\begin{align}\nonumber
a_1&+A_1(a_2-1)+...+A_1A_2...A_{P-1}(a_P-1)\\
&=b_1+B_1(b_2-1)+...+B_1B_2...B_{Q-1}(b_Q-1).
\label{reshape}
\end{align}
The above relationship implicitly assumes column-major order of the corresponding transformation.

{Superscripts in $\{\cdot\}^*$, $\{\cdot\}^\mathrm{T}$, $\{\cdot\}^\mathrm{H}$, and $\{\cdot\}^\dagger$ stand respectively for the complex-conjugate, transpose, Hermitian, and pseudo-inverse of a vector/matrix.
Notations $\mathrm{diag\{\cdot\}}$ and $\left\| \cdot\right\|_F$ represent the diagonalization and Frobenius norm of a matrix.
Symbols $\circ$, $\otimes$, and $\odot$ denote the outer product, Kronecker product and Khatri-Rao product, respectively.
Finally, $\mathrm{Re\{\cdot\}}$ and $\mathrm{Im\{\cdot\}}$ stand for the real and imaginary parts of a complex argument, respectively.

\subsection{Paper Organization}
The reminder of this paper is organized as follows.
Section~\ref{review} reviews the related work and gives motivations.
The multidimensional signal (data collection) model of FMCW MIMO radar is introduced in Section~\ref{secDatamodel}.
Main theoretical results are presented in Sections~\ref{secDenoise} and \ref{secParaEst}. Particularly, the tensor signal denoising via TT is presented in Section~\ref{secDenoise}. Some useful novel theorems about equivalence between CPD and TT formats that are instrumental for developing the proposed algorithm are also presented, while the proofs are given in the Appendices.
ESPRIT-based signal parameter estimation algorithm is presented then in Section~\ref{secParaEst}.
The performance of the proposed method is demonstrated in terms of simulation results and also in terms of an experiment for practical measurements in Section~\ref{Result}.
Section \ref{conclusion} concludes the paper.

\begin{figure}[!t]
\centering
\vspace{-0em}
  {       \includegraphics[width=0.7\columnwidth,draft=false]{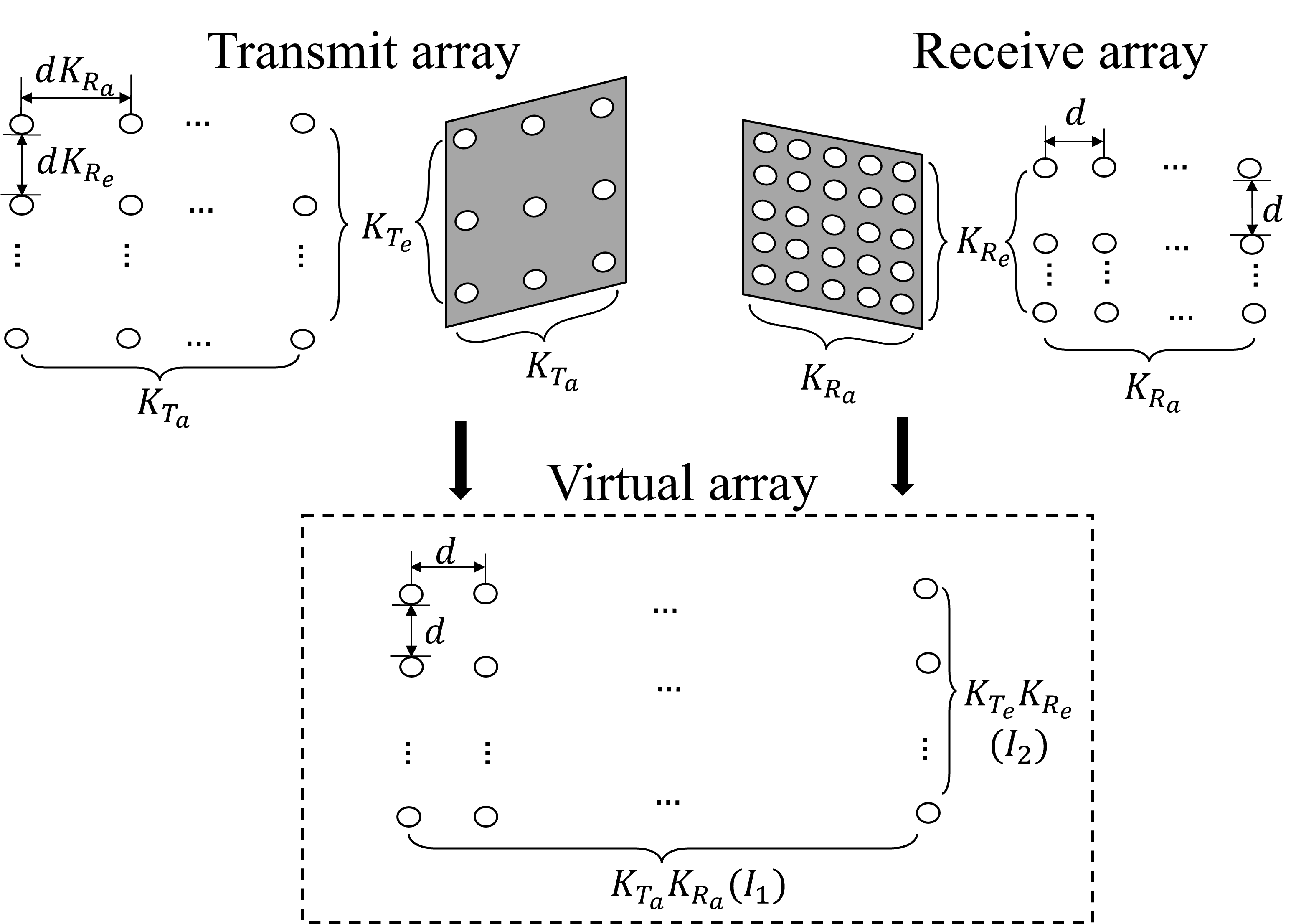}}
\caption{\small The TDM-MIMO transmit and receive arrays.\label{VA}}
  \vspace{-2em}
\end{figure}

\section{Related Work and Motivation}\label{review}
Over the past decades, numerous powerful approaches have been developed for detecting and classifying signals for different radar types.
Traditional signal detection techniques rely on FFT because of its low complexity \cite{9638345,7029030,9903348}.
Nevertheless, the inherent properties of FFT-based approaches, such as side-lobe leakage, low peak power, and wide occupied spectrum, pose challenges for high-resolution sensing applications \cite{SBRA021E}.
 However, they are found to be less attractive especially for the follow-up direction-of-arrival (DOA) estimation without sacrificing angular accuracy \cite{zheng2023coarraytensor}.

To meet the high requirements in angular parameter estimation, subspace-based techniques such as MUSIC and ESPRIT address angular resolution limitations by leveraging spatial spectrum analysis.
While these methods surpass the Rayleigh limit, they inherently reshape multidimensional radar signals into matrices, discarding critical correlations across range, velocity, and angle dimensions.
For instance, MUSIC requires exhaustive 2D spectral searches but cannot jointly estimate all three parameters, failing in multi-target scenarios with overlapping trajectories \cite{7938435}.
While ESPRIT leverages rotational invariance to reduce computational load, it remains confined to matrix-based processing and is unable to exploit higher-order signal structures \cite{485927}.
 Its performance further deteriorates under low SNR conditions.
 Moreover, it has been observed experimentally that its classification performance struggles to achieve high accuracy in low SNR regime, which is a common challenge in FMCW radar systems due to electromagnetic interference and multi-path reflections~ \cite{6361474,10705685}.
This limitation arises as the presence of noise which deteriorates the intrinsic signal characteristics, and thus, is crucial for accurate signal processing.
The noisy target signatures further affect the parameter estimation and feature extraction.

 To improve the signal quality and mitigate the effects of noise, several methods have been developed to reduce noise in signals by analytically leveraging the signal characteristics.
 Recently, deep learning techniques have attempted to mitigate noise through data-driven denoising, but they require large labeled datasets \cite{8481708, 10229208}.
 However, there are still limitations such as lack of large training datasets \cite{9310710}, low efficiency and the need for some practically demanding prior assumptions~\cite{9930810}.
 Hu et al. \cite{9853513} have proposed a time-frequency image denoising method based on neural networks which can contribute to identifying radar waveforms with low SNR.
 Kim et al. \cite{10436154} have designed a patch-based noise reduction framework to enhance the performance of waveform classification based on low probability of intercept (LPI) radar.
 Du et al. \cite{7160673} have proposed to denoise the returned micro-Doppler radar signals via reconstructing the echo within the subspace that is composed of selected principal components and discarding the residual noise subspace under low SNR conditions.
 It can be observed that most denoising frameworks prioritize pulsed or SAR systems, while overlooking FMCW radar's unique modulation characteristics.
 Principal component analysis (PCA)-based methods \cite{10164024} isolate noise subspace but struggle with rank-deficiency of FMCW signals caused by coherent targets or sparse antenna arrays.
 Beyond that, the above mentioned methods do not preserve the multidimensional correlations of higher-order signals essential for joint parameter estimation.

Tensor, which is an extension of matrix to higher-dimensional spaces, exhibits intrinsic multi-linear correlations, hence partially addresses higher-order signal processing for multidimensional FMCW radar signals \cite{zheng2023coarraytensor}.
Consequently, tensor-based methods have emerged as a powerful tool for large-scale signal analysis \cite{7891546, 10284994}.
Among such methods, CPD and Tucker decomposition are widely used to characterize underlying multidimensional parameter structure.
However, CPD reliance on unstable alternate least squares (ALS) iterations \cite{kolda2009tensor}.
 Also Tucker's dimensionality collapses under limited number of antennas \cite{4545266}.
  It renders CPD and Tucker decomposition impractical for real-time FMCW applications.
Both methods also exhibit prohibitive computational cost when scaling to high-dimensional parameter spaces and suffer from difficulties in finding the optimal solutions \cite{10750039}.

As an alternative, TT decomposition has been developed as a resolution to these limitations \cite{oseledets2011tensor}.
Its superiority in characterizing the low-rankness and reducing the computational complexity in various applications, such as hyperspectral image denoising \cite{gong2020tensor}, image recovery \cite{9078764}, channel estimation \cite{10048567}, DOA estimation \cite{9872046}, etc., has been demonstrated.
 By factorizing multidimensional FMCW signals into cascaded low-rank cores, TT preserves couplings across range, velocity, and angle dimensions while suppressing noise through intrinsic subspace separation.
 Unlike MUSIC and ESPRIT, TT avoids matrix flattening, enabling joint parameter estimation without sacrificing spatial resolution.
 Its hierarchical core structure also mitigates rank deficiency caused by coherent signals, which gives critical advantages over PCA and subspace methods.
 Furthermore, TT's computational efficiency and adaptability to limited hardware configurations (e.g., sparse antenna arrays) make TT a tailored solution for higher-order signals in FMCW radar.

 While TT decomposition offers theoretical advantages for multidimensional signal processing, its direct application to FMCW MIMO radar signals faces challenges.
 Firstly, rank selection in conventional TT decomposition is based on prior noise-dependent knowledge and often fails to capture the adaptive multi-scale structures in radar signals.
 In turn, decomposition errors existing in TT implementations would affect the accuracy of parameter estimation, degrading the accuracy of targets' localization.
 To address these limitations, we propose a tensor decomposition-based signal processing framework that overcomes the error accumulation.
  It is then embedded in the noise reduction and parameter estimation pipelines for FMCW MIMO radar signals.
\section{Data Model for Multidimensional Signal Retrieval}\label{secDatamodel}
We consider a monostatic MIMO system that operates at mmWave frequencies and uses multiple antenna planar arrays with $K_T=K_{T_a}K_{T_e}$ collocated transmit and $K_R=K_{R_a}K_{R_e}$ collocated receive antenna elements as shown in Fig.~ \ref{VA}.
To simplify the practical implementation, we use time-division multiplexing (TDM) technique for the virtual antenna configuration shown in Fig.~ \ref{VA}.
 In TDM-based MIMO radar, the transmit antennas emit signals in a time-sequenced manner, where each antenna transmits its waveform in a distinct time slot.
 This sequential transmission ensures orthogonality in the time domain, allowing the receiver to separate and process the reflected signals from different transmit antennas without interference.
The key advantage of TDM in MIMO radar is its ability to achieve high spatial resolution and target detection accuracy while maintaining a relatively simple hardware implementation \cite{8645667}.
By avoiding simultaneous transmission, TDM eliminates cross-talk between antenna elements, which is particularly beneficial in automotive and intelligent transportation applications where real-time processing and reliability are critical.

In Fig.~ \ref{VA}, both the transmit and receive antenna arrays are uniform rectangular arrays (URAs).
At the receiver, the spacings between elements are half the working wavelength, i.e., $d=\lambda/2$, along both azimuth and elevation dimensions.
At the transmitter, the spacings between transmit elements are $d K_{R_a}$ and $d K_{R_e}$ to realize the MIMO virtual array (VA) along azimuth and elevation dimensions, respectively.
Hence, an improved angular resolution can be obtained by the synthesized URA with a smaller number of physical antenna elements ($K_T + K_R$) than the number of virtual elements ($K_T \times K_R$).
To simplify notations, we use $I_1\triangleq K_{Ta}K_{Ra}$ and $I_2\triangleq K_{Te}K_{Re}$ in the following equations, and $i_1\in [1,K_{Ta}K_{Ra}], i_2\in [1,K_{Te}K_{Re}]$ are the indices of virtual elements along azimuth and elevation dimensions.
\begin{figure}[!t]
\centering
\vspace{-0em}
  {       \includegraphics[width=0.9\columnwidth,draft=false]{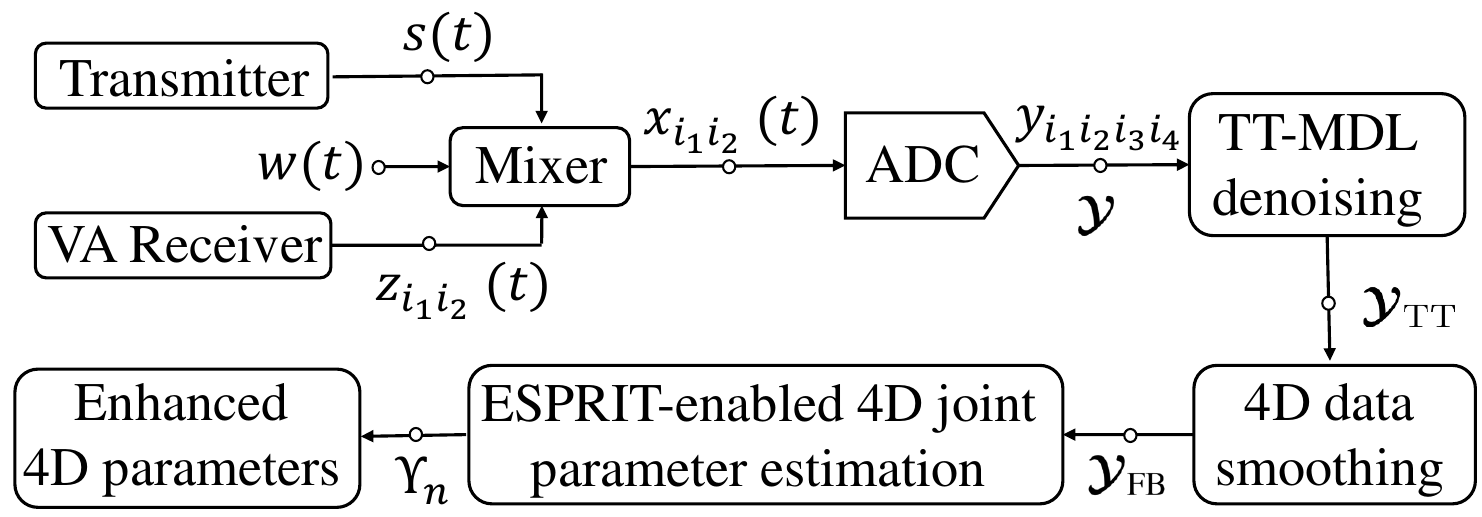}      }
         \vspace{-0.0em}
\caption{\small Block diagram of the proposed method in an FMCW MIMO radar.\label{BlockDiagram}}
  \vspace{-2em}
\end{figure}

The FMCW modulation is widely adopted for mmWave automotive applications of radar systems.
 It is because FMCW allows for simultaneous range-velocity-angle estimation \cite{zhu2022efficient}.
In addition, a wide variety of solid-state transmitters allows for readily compatible modulation, which is commercialized recently with utilization of the Integrated Circuit (IC) technology \cite{lien2016soli}.
Generally, an FMCW radar system transmits periodic linear frequency modulated electromagnetic waves and provides range-velocity-angle estimation with a high resolution.
The resolution is also flexible thanks to the adjustable system bandwidth and chirp duration.
Thus, we utilize here the FMCW with up-chirp modulation for automotive sensing.

The signal $s(t)$ in the $n$-th chirp is sent from each antenna element, that is expressed mathematically using the complex representation as
\begin{equation}
s(t)= e^{  j (2\pi f_c t+ \pi \Delta f t^2)+\phi_0 }, (n-1)T\leq t\leq nT,
\label{eqTX}
\end{equation}
where $f_c$,  $\phi_0$, and $\Delta f\triangleq\frac{B}{T}$ are the starting frequency, initial phase, and the chirp slope rate with $B$ and $T$ denoting the transmitted signal bandwidth and the chirp duration.
Assuming that $R$ targets are reflecting radar signal, that results in a superposition of $R$ delayed versions lagged behind $s(t)$, the signal received by the $(i_1, i_2)$-th virtual antenna element can be expressed as
\begin{align}\nonumber\label{eqRX}
z_{i_1i_2}(t)
=\sum _{r=1}^{R}&\beta_r e^{ j (2\pi f_c (t-\tau_{i_1,i_2,n,r})+ \pi \Delta f (t-\tau_{i_1,i_2,n,r})^2)+\phi_0}
,\\ &(n-1)T\leq t\leq nT,
 \end{align}
where $\beta_r$ is the scaling amplitude backscattered from the $r$-th target.
 After the down-conversion mixer, the intermediate frequency (IF) signal is given by
 \begin{align}\nonumber
x_{i_1i_2}(t)=s^*(t)&z_{i_1i_2}(t)=\sum _{r=1}^{R}\alpha'_{r} e^{j2\pi (\Delta f\tau_{i_1,i_2,n,r} t+f_c\tau_{i_1,i_2,n,r})}, \\
&(n-1)T\leq t\leq nT.\label{eqIF}
 \end{align}
Herein, $\tau_{i_1,i_2,n,r}$ is the round trip delay of $r$-th target's backscattered signal in the $n$-th chirp received by the $(i_1,i_2)$-th virtual antenna element, which is given by
\begin{align}\nonumber
\tau_{i_1,i_2,n,r} &= \frac{2(R_r+v_r nT)}{c}-d(i_1-1)\cos(\phi_r)\sin(\theta_r)\\
&-d(i_2-1)\sin(\phi_r)\sin(\theta_r),
\end{align}
 where $R_r$, $v_r$, $\phi_r$, and $\theta_r$ denote the distance, the velocity, the azimuth, and the elevation of the $r$-th target, respectively.
The constant $\alpha'_r$ is the amplitude of the IF signal of the $r$-th target.

As shown in Fig. \ref{BlockDiagram}, the IF signals $x_{i_1,i_2}(t)$, $i_1\in[1,I_1], i_2\in[1,I_2]$ go to the analog-to-digital converter (ADC) for discrete sampling.
Herein, denote the sampling interval as $T_s$, and the discrete point in time at the $i_3$-th sample of the $n$-th chirp as $t=i_3T_s+(n-1)T$.
  For consistency in expression, the index $i_4$ is used referring to the index $n$.
By inserting the time notation into equation (\ref{eqIF}), the output of ADC is given as
 \begin{equation}
y_{i_1,i_2,i_3,i_4} \triangleq x_{i_1,i_2}(i_3T_s+(i_4-1)T) =\sum _{r=1}^{R}\alpha_{r} \prod_{m=1}^4 \boldsymbol{a}_{r,I_m}(i_m), \label{eqIFDiscrete}
\end{equation}
where $\boldsymbol{a}_{r,I_m}(i_m)$ denotes the $i_m$-th element of the vector $\boldsymbol{a}_{r,I_m}$.
Herein, $\boldsymbol{a}_{r,I_1}$ and $\boldsymbol{a}_{r,I_2}$ are the steering vectors of URAs, that are expressed as
 \begin{align}\label{a1}
 \boldsymbol{a}_{r,I_1}=[1, e^{-j2\pi\Theta_r },..., e^{-j2\pi\Theta_r(I_1-1) }]^{\mathrm{T}},\\\label{a2}
 \boldsymbol{a}_{r,I_2}=[1, e^{-j2\pi \Phi_r},..., e^{-j2\pi \Phi_r(I_2-1)}]^{\mathrm{T}},
\end{align}
where $\Theta_r\triangleq f_c d\cos(\phi_r)\sin(\theta_r)$ and $\Phi_r\triangleq f_c d\sin(\phi_r)\sin(\theta_r)$ are defined as the spatial frequencies of the $r$-th target.
Similarly, $\boldsymbol{a}_{r,I_3}$ and $\boldsymbol{a}_{r,I_4}$ can be expressed as
 \begin{align}\label{a3}
 \boldsymbol{a}_{r,I_3}&=[e^{-j2\pi\eta_r }, e^{-j2\pi2\eta_r },..., e^{-j2\pi\eta_rI_3 }]^{\mathrm{T}},\\\label{a4}
 \boldsymbol{a}_{r,I_4}&=[1, e^{-j2\pi\mu_r },..., e^{-j2\pi\mu_r(I_4-1) }]^{\mathrm{T}},
\end{align}
where $\eta_r\triangleq 2\Delta f R_r T_s/c$ and
$\mu_r\triangleq v_r T$/ $\lambda$.

Let us now introduce the CPD model and the relevant properties for the signal modeling.
The CPD factorizes an $N$-order tensor $\boldsymbol{\mathcal{X}}_{\mathrm{CP}}\in \mathbb{C}^{I_1 \times I_2...\times I_N}$ into a sum of $R$ rank-$1$ tensors, which can be expressed as
 \begin{align} \nonumber
\boldsymbol{\mathcal{X}}_{\mathrm{CP}}&=\sum_{r=1}^{R}\alpha_r \boldsymbol{u}_{1,r}\circ \boldsymbol{u}_{2,r}\circ...\circ\boldsymbol{u}_{N,r} \\
&=\boldsymbol{\mathcal{A}}\times_1 \boldsymbol{U}_1\times_2 \boldsymbol{U}_2 ... \times_N \boldsymbol{U}_N =[\![\boldsymbol{\alpha};\boldsymbol{U}_1, \boldsymbol{U}_2,...,\boldsymbol{U}_N]\!], \label{CPD}
\end{align}
where $\boldsymbol{\alpha}=[\alpha_1,...,\alpha_R]$ is the weight vector, $\boldsymbol{\mathcal{A}}$ is the $N$-order tensor with the $r$-th diagonal element being ${\alpha}_r$.
 The smallest value of $R$ is called the CP rank, and the $n$-th factor matrix of size $I_n \times R$ along the $n$-th mode is denoted as $\boldsymbol{U}_n=[\boldsymbol{u}_{n,1},..., \boldsymbol{u}_{n,R}]$.

Define the CPD mode-$n$ unfolding of a tensor $\boldsymbol{\mathcal{X}}$ as $\mathrm{unfold}_{\langle n \rangle}(\boldsymbol{\mathcal{X}})=\boldsymbol{X}_{\langle n \rangle} \in \mathbb{C}^{\prod_{l=1}^{n}I_l\times \prod_{l={n+1}}^{N}I_l}$.
The following property is used in this work
\begin{equation}
\boldsymbol{X}_{\langle n \rangle}=(\boldsymbol{U}_{n}\odot...\odot\boldsymbol{U}_{1})\mathrm{diag}\boldsymbol{(\alpha)}(\boldsymbol{U}_N\odot...\odot\boldsymbol{U}_{n+1})^{\mathrm{T}}.
\end{equation}
The inverse operator to unfolding is denoted as $\boldsymbol{\mathcal{X}}=\mathrm{fold}_{\langle n \rangle}(\boldsymbol{X}_{\langle n \rangle})$.
Correspondingly, the mode-$n$ unfolding operator on a tensor is denoted as $\mathrm{unfold}_{(n)}(\boldsymbol{\mathcal{X}}) =\boldsymbol{X}_{(n)}\in \mathbb{C}^{I_n\times I_{n+1}...I_NI_1...I_{n-1}}$.

Furthermore, the following properties hold true:
  \begin{align}\nonumber
  (\boldsymbol{\mathcal{X}}\times_1 \boldsymbol{S}_1&\times_2 \boldsymbol{S}_2 ... \times_N \boldsymbol{S}_N)_{\langle n \rangle}\\
=&(\boldsymbol{S}_{n}\otimes...\otimes\boldsymbol{S}_{1})\boldsymbol{X}_{\langle n \rangle}(\boldsymbol{S}_N\otimes...\otimes\boldsymbol{S}_{n+1})^{\mathrm{T}}.
\end{align}
  \begin{align}\nonumber
  (\boldsymbol{\mathcal{X}}\times_1 \boldsymbol{S}_1&\times_2 \boldsymbol{S}_2 ... \times_N \boldsymbol{S}_N)_{(n)}\\
=&\boldsymbol{S}_{n}\boldsymbol{X}_{(n)}(\boldsymbol{S}_{n+1}\otimes...\otimes\boldsymbol{S}_{N}\boldsymbol{S}_{1}\otimes...\otimes\boldsymbol{S}_{n-1})^{\mathrm{T}}.
\end{align}

According to (\ref{eqIFDiscrete}) and the CPD model in (\ref{CPD}), the low-rank IF signal, that is to be denoised, can be expressed as
 \begin{align}
\boldsymbol{\mathcal{Y}}=[\![\boldsymbol{\alpha};\boldsymbol{U}_1, \boldsymbol{U}_2, \boldsymbol{U}_3, \boldsymbol{U}_4]\!] \in \mathbb{C}^{I_1 \times I_2 \times I_3 \times I_4},
\label{problem}
\end{align}
where $\boldsymbol{U}_n=[\boldsymbol{a}_{1,I_n},...,\boldsymbol{a}_{R,I_n}]\in \mathbb{C}^{I_n\times R}, n=1,2,3,4$.
Herein, we introduce the Vandermonde matrix which accounts for the use of FMCW modulation and URA structure.
Given the factor matrix $\boldsymbol{U}\in \mathbb{C}^{I\times R}$, it is said to be Vandermonde if it can be expressed as
\begin{equation}
\boldsymbol{U}=[\boldsymbol{a}_{1,I},...,\boldsymbol{a}_{R,I}]\in \mathbb{C}^{I\times R}
\end{equation}
where the vectors $\boldsymbol{a}_{1,r} = [1,q_r,q_r^2,...,q_r^{I-1}]^{\mathrm{T}}\in \mathbb{C}^{I\times 1}, r=1...,R$, are generating vectors for $\boldsymbol{U}$ that obey the Vandermonde structure.
Hence, combining the equations (\ref{a1})-(\ref{a4}), $\boldsymbol{U}_n$, $n=1,2,3,4$, can be seen to be Vandermonde matrices.
Properties of Vandermonde matrices will be used for parameter estimation in Section \ref{secParaEst}.

Generally, the observed signal tensor, denoted as $\hat{\boldsymbol{\mathcal{Y}}}$, is contaminated by a mixture of different kinds of noise terms.
 Overall, the noise is assumed to be Gaussian involving amplitude noise and phase noise.
The noise term introduces random fluctuations in the IF signal envelope and jitter in measured beat frequency.
 This leads to inaccuracies directly related to the target range, where the higher the phase noise, the greater the uncertainty in range estimation.
 Besides, since velocity estimation often uses the change of IF signals over time, the phase jitter would translate into velocity measurement errors.
For angle estimation in multiple-antenna FMCW systems, these noise affects the phase information, leading to errors in finding the directions of arrival (DOAs) for the signals.
In this paper, all these factors are combined in a noise tensor $\boldsymbol{\mathcal{W}}$ with $w_{i_1,i_2,i_3,i_4}$ denoting the $(i_1,i_2,i_3,i_4)$-th element in the noise tensor.

Given the noisy signal $\hat{\boldsymbol{\mathcal{Y}}}$ with additive noise $\boldsymbol{\mathcal{W}}$, we aim to reduce the noise and estimate the information for all targets along four dimensions, which are the distance $R_r$, velocity $v_r$, azimuth angle $\phi_r$, and elevation angle $\theta_r$ of the $r$-th target.
Then, the to-be-estimated $\hat{\boldsymbol{\mathcal{Y}}}$ is modeled as
 \begin{align}
\boldsymbol{\mathcal{Y}}=\hat{\boldsymbol{\mathcal{Y}}}+\boldsymbol{\mathcal{W}}.\label{problem}
\end{align}

Equipped with the above mentioned multidimensional signal model, we proceed to address the main problem of noise reduction for a more accurate parameter estimation.

\section{Noise Reduction with Tensor Train Decomposition}\label{secDenoise}
The objective of the noise reduction is to estimate $\hat{\boldsymbol{\mathcal{Y}}}$ from the observed ${\boldsymbol{\mathcal{Y}}}$ by exploring the structures that have to be present in the clean $\hat{\boldsymbol{\mathcal{Y}}}$.
As (\ref{eqIFDiscrete}) shows, the received signals are composed of the target echoes, which are sparse.
Thus, the undetermined problem of denoising can be addressed by enforcing sparsity of $\boldsymbol{\mathcal{Y}}$.
Accordingly, only few singular values are dominant and related to the low rankness of $\boldsymbol{\mathcal{Y}}$, where the singular value decomposition (SVD) is performed on the mode-1, mode-2, mode-3, and mode-4 unfolding matrices $\boldsymbol{Y}_{(1)}$, $\boldsymbol{Y}_{(2)}$, $\boldsymbol{Y}_{(3)}$, $\boldsymbol{Y}_{(4)}$.
To exploit the tensor low-rankness together with its higher order, we next introduce the TT decomposition and also prove its equivalence to CPD.

\subsection{Equivalence Between CPD and TT Decomposition}
 TT decomposition is used to effectively characterize the low-rank structure of higher-order signal tensor and achieve higher storage efficiency.
It decomposes an $N$-order tensor $\boldsymbol{\mathcal{X}}_{\textrm{TT}}$ into a circular multilinear product over $N$ three-order core tensors $\boldsymbol{\mathcal{G}}_n\in \mathbb{C}^{T_{n-1}\times I_{n}\times T_{n}}$, $n=1,2,3$ as
\begin{equation}
\boldsymbol{\mathcal{X}}_{\textrm{TT}}=\boldsymbol{{G}}_1\times^1_2 \boldsymbol{\mathcal{G}}_2 \times^1_3 \boldsymbol{\mathcal{G}}_3 ... \times^1_{N-1} \boldsymbol{\mathcal{G}}_{N-1} \times^1_N \boldsymbol{{G}}_N,
\label{TT}
\end{equation}
where the $n$-th TT cores are $\boldsymbol{\mathcal{G}}_n \in \mathbb{C}^{T_{n-1} \times I_n\times T_n}$, for $1<n<N$, the head $\boldsymbol{{G}}_1\in \mathbb{C}^{I_1\times T_1}$, for $n=1$, and the tail $\boldsymbol{{G}}_N\in \mathbb{C}^{T_{N-1}\times I_N}$, for $n=N$.
Herein, $[T_1, T_2,..., T_{N-1}]$ are called TT ranks.
In the elementwise format, we have
\begin{equation}
\boldsymbol{\mathcal{X}}_{\textrm{TT}}(i_1,...,i_n,...,i_N)=\boldsymbol{{G}}_1(i_1,\cdot)...\boldsymbol{\mathcal{G}}_n(i_n)...\boldsymbol{{G}}_N(\cdot,i_N),
\label{TTDele}
\end{equation}
where $\boldsymbol{\mathcal{G}}_n(i_n)$ denotes the $i_n$-th lateral slice matrix of the core tensor $\boldsymbol{\mathcal{G}}_n, n=1,2,..., N$,
${\boldsymbol{G}}_1(i_1,\cdot)$ denotes $i_1$-th column of $\boldsymbol{G}_1$, and ${\boldsymbol{G}}_N(\cdot,i_N)$ denotes $i_N$-th row of $\boldsymbol{G}_N$.
CPD is essentially a generalized form of Tucker decomposition for a uniform rank.
The algebraic equivalence between the Tucker decomposition and the TT decomposition is established recently in \cite{zniyed2020high}, which inspires the interest in deriving the relationship between the CPD and TT decomposition.

The following theorems establish the relationship between the CPD and TT decomposition.

\emph{{Theorem 1}}:
Assume that the CPD tensor $\boldsymbol{\mathcal{X}}_{\mathrm{CP}}\in \mathbb{C}^{I_1 \times I_2...\times I_N}$ is decomposed as in (\ref{CPD}), then, a TT decomposition of $\boldsymbol{\mathcal{X}}_{\textrm{TT}}$ is given by
\begin{align}\nonumber
\boldsymbol{{G}}_1=& \; \boldsymbol{{U}}_1,\quad {\rm and} \quad  \boldsymbol{{G}}_N=\boldsymbol{{U}}_N^{\mathrm{T}}, \\ \nonumber
\boldsymbol{\mathcal{G}}_n=& \; \boldsymbol{\mathcal{T}}_n\times_2 \boldsymbol{{U}}_n, 1<n<\bar{n},\\\nonumber
                           &\textrm{with}~\boldsymbol{\mathcal{T}}_n=\mathrm{reshape}\left(\boldsymbol{I}_{R^n},\left[R^{n-1},R,R^{n}\right] \right),\\\nonumber
\boldsymbol{\mathcal{G}}_{\bar{n}}=& \; \boldsymbol{\mathcal{T}}_{\bar{n}}\times_2 \boldsymbol{{U}}_{\bar{n}}, \\\nonumber
                           &\textrm{with}~\boldsymbol{\mathcal{T}}_{\bar{n}}=\mathrm{reshape}\left(\boldsymbol{\mathcal{A}},\left[R^{\bar{n}-1},R,R^{N-\bar{n}}\right] \right),\\\nonumber
\boldsymbol{\mathcal{G}}_n=& \; \bar{\boldsymbol{\mathcal{T}}}_{n}\times_2 \boldsymbol{{U}}_n, \bar{n}<n<N,\\\nonumber
                           &\textrm{with}~\bar{\boldsymbol{\mathcal{T}}}_{n}=\mathrm{reshape}\left(\boldsymbol{I}_{R^{N-n+1}},\left[R^{N-n+1},R,R^{N-n}\right] \right),
\end{align}
where $\bar{n}$ is the smallest $n$ that verifies $R^n\geq R^{N-n}$, i.e., $n=\lceil\frac{N}{2}\rceil$, and the TT ranks verify that $T_n=\min (R^n,R^{N-n})$.
The operator $\mathrm{reshape}$ is defined by the relationship in (\ref{reshape}).

The proof of \emph{Theorem} 1 can be found in Appendix A.
\label{CPD2TT}

\emph{Theorem} 1 stresses on the fact that by computing a CPD via its associated TT decomposition, the model structure is maintained, while the complexity is significantly reduced as it will be discussed later.
However, there still remains an inequivalence between the ranks of CPD and TT decomposition, which is resolved by another theorem that appears next.
It is worth mentioning that \emph{Theorem}~1 presents the necessary but not sufficient condition for obtaining the smallest of TT ranks, which means that the theorem remains true even if some factor matrices $\boldsymbol{U}_n$ in (\ref{CPD}) are rank deficient.
Specifically, the algebraic equivalence does not guarantee the smallest value of TT ranks and the estimated TT decomposition can vary accordingly.

The following theorem gives the relationship between the TT rank and CPD rank,
where the lower bound of TT ranks is related to the multi-linear structure of the tensor.

\emph{Theorem 2}:
Assume that all parameters of $\boldsymbol{U}_n$ are  continuous with respect to the Lebesgue measure in $\mathbb{C}^{R}$.
 For a CPD tensor $\boldsymbol{\mathcal{X}}_{\mathrm{CP}}$ satisfying (\ref{CPD}), the smallest TT ranks leading to a minimum $\|\boldsymbol{\mathcal{X}}_{\mathrm{CP}}-\boldsymbol{\mathcal{X}}_{\mathrm{TT}}\|_\mathrm{F}^2$ satisfy
\begin{equation}
R_n=\min\left( \prod_{l=1}^{n}I_l,R,\prod_{l=n}^{N}I_l \right), ~n=1,2,..., N.
\end{equation}
The proof of \emph{Theorem} 2 can be found in Appendix B.

\subsection{Improved TT Denoising Implementation}
As follows from \emph{Theorem} 2, TT decomposition actually employs low-rank constrains on the tensor $\boldsymbol{\mathcal{X}}_{\mathrm{CP}}$.
Furthermore, according to Proposition 1 and Corollary 1 in \cite{gong2020tensor}, a tensor with a low TT rank may fail to possess a low CP rank, which especially occurs when there exist targets with the same parameters in a certain dimension.
Thus, by adopting CPD model, we cannot exploit the tensor's low-rankness as much as we can do it using the TT model that comes with the potential advantage of modeling the latent low-rankness.

Particularly, we employ the TT-SVD decomposition using the MDL method to characterize the sparsity and correlation among different dimensions in the signal tensor ${\boldsymbol{\mathcal{Y}}}$, which we refer to as \textit{TT-MDL} in the following.

Mathematically, the noise reduction problem consists of solving the following tensor low-rank approximation problem
\begin{equation}\label{optimazation1}
\underset{\boldsymbol{\mathcal{Y}}_{\mathrm{TT}}}{\mathrm{min}} \left\| {\boldsymbol{\mathcal{Y}}}-\boldsymbol{\mathcal{Y}}_{\mathrm{TT}} \right\|_F^2, \;
\mathrm{s.t.} \;\boldsymbol{\mathcal{Y}}_{\mathrm{TT}} =\boldsymbol{\mathcal{G}}_1\times^1_2 \boldsymbol{\mathcal{G}}_2 \times^1_3 \boldsymbol{\mathcal{G}}_3 \times^1_4 \boldsymbol{\mathcal{G}}_4,
\end{equation}
where the decomposed cores are $\boldsymbol{\mathcal{G}}_n \in \mathbb{C}^{T_{n-1} \times I_n\times T_n}$, here $n=1,2,3,4$, and $T_0=T_4=1$.
We denote $\boldsymbol{\mathcal{G}}_1\in \mathbb{C}^{1 \times I_1\times T_1}$ and $\boldsymbol{\mathcal{G}}_4\in \mathbb{C}^{T_{3}\times I_4 \times 1}$ as respective matrices.
Herein, we find the TT decomposition using $N-1$ sequential SVDs of auxiliary matrices found via TT-SVD algorithm \cite{oseledets2011tensor}.

Proceeding by induction of an iterative variable with the initial value of $\boldsymbol{C}=\boldsymbol{Y}_{(1)}$, the $T_n$-truncated SVD of $\boldsymbol{C}$ is given by
\begin{equation}\label{mode1SVD}
\boldsymbol{C}=\boldsymbol{U}_{1}\boldsymbol{\Sigma}_{1}\boldsymbol{V}^{\mathrm{T}}_{1}+\boldsymbol{E}_{(1)}
\!=\!\sum_{t_1=1}^{T_1}\boldsymbol{U}_{1}(\cdot,t_1)\boldsymbol{C}_0(t_1,\cdot) + \boldsymbol{E}_{(1)}.
\end{equation}

Similar to $\boldsymbol{C}$, $\boldsymbol{E}_{(1)}$ denotes the noise components corresponding to the mode-$1$ unfolding of noise tensor $\boldsymbol{\mathcal{W}}$.
Using SVD, we obtain the unitary matrices $\boldsymbol{U}_{1}$ and $\boldsymbol{V}_{1}$ consisting of eigen-vectors, and the diagonal matrix $\boldsymbol{\Sigma}_{1}$ with its diagonal elements being the first $T_{n}$ eigenvalues ordered in descending order.
Moreover, auxiliary variable $\boldsymbol{C}_0 =\boldsymbol{\Sigma}_{n}\boldsymbol{V}^{\mathrm{T}}_{n}\in \mathbb{C}^{T_{n-1} \times \prod_{m=n}^{4} I_m}$ is introduced to update $\boldsymbol{C}$ in each iteration except the last one, which is given as
\begin{align}
\boldsymbol{C}=\mathrm{reshape}\left(\boldsymbol{C}_0,\left[T_{n-1}I_n, \prod_{m=n+1}^{4} I_m\right]\right).\label{reshapeC}
\end{align}

Then, the next step is to perform the $T_n$-truncated SVD on $\boldsymbol{C}$, which can be written as
\begin{align}\label{mode1SVD2}
\boldsymbol{C}=\boldsymbol{U}_{n}\boldsymbol{\Sigma}_{n}\boldsymbol{V}^{\mathrm{T}}_{n}+\boldsymbol{E}_{n}.
\end{align}
After finding $\boldsymbol{U}_{n}$, $\boldsymbol{\mathcal{G}}_n$ is obtained by the following reshaping
\begin{align}
\boldsymbol{\mathcal{G}}_n=\mathrm{reshape}(\boldsymbol{U}_n, [T_{n-1},I_n,T_n]).\label{reshapeG}
\end{align}

Note that the TT-SVD algorithm in \cite{oseledets2011tensor} requires the prescribed accuracy as an input parameter to predetermine TT ranks for truncated SVD.
 The prescribed accuracy in turns refers to the noise variance of the input signal $\boldsymbol{\mathcal{Y}}_{\mathrm{TT}}$ as a limit of such accuracy.
However, it is impractical to obtain the noise variance based on the received signal, as well as the TT ranks.
Thus, the MDL \cite{MDL1985} information theoretic criteria is adopted here to estimate the TT ranks.
  MDL is commonly used to determine the number of target signals without the need of subjective threshold.
By integrating MDL, the model complexity and low-rank structure accuracy are balanced, which robustly adapts to unknown noise levels while preserving critical signal features.

Specifically, $\boldsymbol{C}$ is first centered as $\overline{\boldsymbol{C}}$ by subtracting the column means.
 The covariance matrix of $\overline{\boldsymbol{C}}$ is then computed as $\boldsymbol{\Delta}=\overline{\boldsymbol{C}}\overline{\boldsymbol{C}}^{\mathrm{H}}/N$ to further employ SVD.
MDL is implemented on $\overline{\boldsymbol{C}}$, rather than $\boldsymbol{Y}_{(n)}$, to estimate the TT rank of the corresponding mode.
 It is done in order to reduce the computational complexity.
Thus, MDL problem is given as
\begin{align}\label{MDL}\nonumber
\mathrm{MDL}({\boldsymbol{C}})=&\arg\underset{\hat{t}}{\min} -\log\left( \frac{\prod_{k=\hat{t}+1}^{M}\lambda_{k}^{1/M}}{\frac{1}{M-t}\sum_{k=\hat{t}+1}^{M}\lambda_k} \right)^{N(M-\hat{t})}\\
&+\frac{1}{2}\hat{t}(2M-\hat{t})\log N ,
\end{align}
where $M\triangleq T_{n-1}I_n$, $N\triangleq \prod_{k=n+1}^{N}I_k$ for the $n$-th mode, and $\lambda_k,~k=1,2,..., M$ are the eigenvalues of $\boldsymbol{\Delta}$.
The detailed algorithm for the proposed method is summarized in Algorithm~1.
\begin{table}[!t]
\label{TTSVD}
\begin{threeparttable}
\begin{tabular}{p{0.9\columnwidth}}
\toprule
\textbf{Algorithm 1:} TT-MDL denoising algorithm \\
\midrule
 \textbf{Input:} The $4$-order tensor $\boldsymbol{\mathcal{Y}}\in \mathbb{C}^{I_1 \times I_2\times I_3 \times I_4}$.\\
 \textbf{Output:} The low-rank approximation $\boldsymbol{\mathcal{Y}}_{\mathrm{TT}}$ of tensor $\boldsymbol{\mathcal{Y}}$ with TT cores $\boldsymbol{\mathcal{G}}_n, n=1,2,3,4$.\\
 \textbf{Initialization:} $N=4$, $T_0=1$, $\boldsymbol{C}=\boldsymbol{Y}_{(1)}$.\\
1. \textbf{for} $n=1:N-1$ \textbf{do}\\
2. Calculate rank $T_n=\mathrm{MDL}({\boldsymbol{C}})$ by solving (\ref{MDL}).\\
3. Calculate the $T_n$-truncated SVD by (\ref{mode1SVD2}).\\
4. Calculate $\boldsymbol{C}_0=\boldsymbol{\Sigma}_{n}\boldsymbol{V}^{\mathrm{T}}_{n}$.\\
5. Update $\boldsymbol{C}$ by (\ref{reshapeC}).\\
6. Update $\boldsymbol{\mathcal{G}}_n$ by (\ref{reshapeG}).\\
7. \textbf{end for }\\
8. Calculate $\boldsymbol{\mathcal{G}}_N=\mathrm{reshape}(\boldsymbol{C}, [T_{N-1},I_N,T_N])$.\\
\textbf{Output:}\;The core tensors $\boldsymbol{\mathcal{G}}_n, n=1,2,3,4$, and $\boldsymbol{\mathcal{Y}}_{\mathrm{TT}}=\boldsymbol{\mathcal{G}}_1\times^1_2 \boldsymbol{\mathcal{G}}_2 \times^1_3 \boldsymbol{\mathcal{G}}_3 \times^1_4 \boldsymbol{\mathcal{G}}_4$.\\
\bottomrule
\end{tabular}
\end{threeparttable}  \vspace{-1em}
\end{table}
\subsection{Discussion}
Computations associated with the multidimensional CPD signal model for directly performing parameter estimation using ALS are high, as it follows from the large number of iterations and the need of good initialization.
Besides, the single global CP-rank $R$ constraint in CPD often inadequately separates subtle low-rank features from high-dimensional noise components which are coupled with signal factors, especially when there exist targets with the same parameters in a certain dimension.
Moreover, determining the CP rank of order $n>2$ or deciding whether a rational tensor has CP-rank is an NP-hard problem.
This brings challenges for further calculating low-rank approximations, and leads to severe difficulties if the canonical format is used.
As an illustration, the best low-rank approximation of an $N$-order tensor $\boldsymbol{\mathcal{X}}_{\mathrm{CP}}\in \mathbb{C}^{I_1 \times I_2...\times I_N}$ with CP-rank $R$ is given by
\begin{equation}
\boldsymbol{\mathcal{X}}^*=\underset{\boldsymbol{\mathcal{Y}}\in \mathbb{C}^{I_1 \times I_2...\times I_N}, \mathrm{CP-rank}(\boldsymbol{\mathcal{Y}})\le R }{\mathrm{argmin}}(\left \|\boldsymbol{\mathcal{X}}-\boldsymbol{\mathcal{Y}}  \right \| ).\label{CPapproximation}
\end{equation}
The selection of the norm $\left \| \cdot\right \| $ depends on a particular application.
 However, it is proved in \cite{CPill} that problem (\ref{CPapproximation}) is ill-posed regardless of the choose of $\left \| \cdot\right \|$.

Unlike global rank constraints in CPD, TT model is more powerful in representing the latent low-rankness of tensors, which is discussed above.
Referring to the equivalent reletionship between CPD and TT in \emph{Theorem 1}, the multi-scale signal features are decoupled and organized into hierarchically ordered subspaces in TT format.
However, the resultant TT ranks are increased and remain suboptimal due to residual noise entangled in redundant subspaces.
To show this suboptimality compared to TT-MDL, a QR decomposition enabled recompression procedure is integrated by hierarchically truncating localized singular values across original TT cores.
  Itdisentangles noise-corrupted dimensions from the intrinsic low-rank signal structure.

Specifically, CPD is firstly applied to obtain $\boldsymbol{\mathcal{X}}_{\mathrm{CP}}$ with an increased global CP-rank and the corresponding product of mode-$n$ unfolding matrices $\boldsymbol{X}_{(n)}$, that is,
\begin{equation}
\boldsymbol{X}_{(n)}=\boldsymbol{\hat{M}}\boldsymbol{\hat{N}}^{\mathrm{T}},
\end{equation}
where
\begin{equation}
\begin{aligned}
&\boldsymbol{\hat{M}}=\boldsymbol{{G}}_1...\times^1_n \boldsymbol{\mathcal{G}}_n\in\mathbb{C}^{I_1...I_n\times \hat{T}_n}, \\ &\boldsymbol{\hat{N}}=\boldsymbol{\mathcal{G}}_{n+1}...\times^1_n \boldsymbol{{G}}_N\in\mathbb{C}^{\hat{T}_{n}\times I_{n+1}...I_N}.
\end{aligned}
\end{equation}
Then, QR decompositions are conducted to orthogonalize $\boldsymbol{\hat{M}}$ and $\boldsymbol{\hat{N}}$ sequentially,
 which enforces left orthogonality for reducing suboptimal rank $\hat{T}_n$, that is,
\begin{equation}
\begin{aligned}
&\boldsymbol{\hat{M}}=\boldsymbol{A}_\mathrm{M}\boldsymbol{B}_\mathrm{M},\\ &\boldsymbol{\hat{N}}=\boldsymbol{A}_\mathrm{N}\boldsymbol{B}_\mathrm{N}.
\end{aligned}
\end{equation}
where $\boldsymbol{A}_\mathrm{M}\in\mathbb{C}^{I_1...I_n\times \hat{T}_n}$ and $\boldsymbol{A}_\mathrm{N}\in\mathbb{C}^{\hat{T}_{n}\times I_{n+1}...I_N}$ are orthogonal matrices that isolate dominant signal components into an orthogonal basis, while $\boldsymbol{B}_\mathrm{M}\in\mathbb{C}^{\hat{T}_n\times \hat{T}_n}$ and $\boldsymbol{B}_\mathrm{N}\in\mathbb{C}^{\hat{T}_n\times \hat{T}_n}$ are upper-triangular matrices that concentrate residual energy including noise.
The localized truncated-SVD is conducted over the reconstructed matrix $\boldsymbol{B}_\mathrm{M}\boldsymbol{B}_\mathrm{N}^{\mathrm{T}}$ as
\begin{equation}
\boldsymbol{B}_\mathrm{M}\boldsymbol{B}_\mathrm{N}^{\mathrm{T}}=\boldsymbol{D}_1\boldsymbol{\Lambda}\boldsymbol{D}_2^{\mathrm{T}},
\end{equation}
where $\boldsymbol{\Lambda}\in\mathbb{C}^{{T}_n\times {T}_n}$ is a diagonal matrix,  $\boldsymbol{D}_1\in\mathbb{C}^{\hat{T}_n\times {T}_n}$ and $\boldsymbol{D}_2\in\mathbb{C}^{\hat{T}_n\times {T}_n}$ are singular matrices with orthogonal columns.
The optimized rank $T_n$ is obtained under the accuracy constraint
\begin{equation}
\|\boldsymbol{\Lambda}-\boldsymbol{\hat{\Lambda}}\|_{\mathrm{F}}\leq \varepsilon\|\boldsymbol{\hat{\Lambda}}\|_{\mathrm{F}},
\end{equation}
where $\varepsilon$ is the accuracy tolerance, and $\hat{\boldsymbol{\Lambda}}$ is obtained by the original non-truncated SVD over $\boldsymbol{B}_\mathrm{M}\boldsymbol{B}_\mathrm{N}^{\mathrm{T}}$. In this way, the recompression is achieved by the optimized rank $T_n$ selection across core tensors.
Unlike global rank constraints in CPD, this hierarchical truncation allows per-core rank adaptation across different tensor modes, showing the improved noise reduction under TT format compared to CPD.

Furthermore, the computational complexity of decomposing TT model is reduced by separating the modes of a tensor into $N$ tensors with orders two or three.
Note that CPD can be regarded as a special format of Tucker decomposition if the core tensor is diagonal.
Hence, the computational benefit can be evaluated in contrast with computing Tucker model by means of SVD, which is referred to as the higher-order singular value decomposition (HOSVD).
The computational complexity of TT-SVD is of order $\mathcal{O}(RI^Q)$ with $R$, $I$, and $Q$ denoting the TT-rank, the largest dimension, and the order of the signal tensor, respectively \cite{zniyed2020high}.
 Meanwhile, the computational complexity of HOSVD is of order $\mathcal{O}(QRI^Q)$, which scales linearly by an additional factor $Q$.
This indicates that the cost of computing a single factor by HOSVD is comparable to the cost of computing $Q$ factors by TT-MDL.

\section{Parameter estimation}\label{secParaEst}
 The conventional ESPRIT utilizes the inherent rotational invariance of a signal, yet follows the principles of matrix-based signal processing for parameter estimation.
However, it shows the potential to exploit the multi-linear signal structure as rotational invariance of signal exists along the multiple dimensions of the signal.
 Thus, we exploit extensions of ESPRIT to higher-order methods for parameter estimation in this section.
Before that, we also deduce the pre-processing data smoothing by employing FBA to enhance the estimation accuracy.
\subsection{Data Smoothing}\label{sec3.3}
When more than one target is present, rank deficiency problem may occur caused by coherent target signals, e.g., targets appearing at the same range from the sensing system, but at different angles or with different velocities.
Thus, before parameter estimation, a data smoothing, including the spatial smoothing FBA as a preprocessing step, is used to mitigate the influence of rank deficiency for enhancing the robustness.

\textbf{Spatial Smoothing (SS):}
The spatial smoothing aims to replace a rank deficient tensor with a full rank tensor.
Note that the spatial factors, i.e., matrices $\boldsymbol{U}_n$, $n=1,2,3,4$, have Vandermonde structure.
 We apply the spatial smoothing along the four modes of the signal tensor $\boldsymbol{\mathcal{Y}}$, which refers to a 4D spatial smoothing.
Specifically, define the SS sub-dimensions as TT rank $T_n=\mathrm{MDL}(\overline{\boldsymbol{C}}), n=1,2,3,4$, and the corresponding SS parameters are then obtained as $L_n=I_n-T_n+1, n=1,2,3,4$, respectively.
Given $I_n$, the selection matrices $\boldsymbol{S}_{l_1,l_2,l_3,l_4}$ are defined as
\begin{align}
\boldsymbol{S}_{l_1,l_2,l_3,l_4}&=\boldsymbol{S}_{\hat{l}_1}\otimes \boldsymbol{S}_{\hat{l}_2} \otimes \boldsymbol{S}_{\hat{l}_3}\otimes \boldsymbol{S}_{\hat{l}_4},
\end{align}
where $\boldsymbol{S}_{\hat{l}_n}=[\boldsymbol{0}_{T_n \times (l_n-1)} \boldsymbol{I}_{T_n} \boldsymbol{0}_{T_n \times (L_n-l_n)}]\in \mathbb{C}^{T_n\times I_n}, n=1,2,3,4$.

Define $R\triangleq \min\{T_n\}$, then the spatial smoothing operation for the signal tensor can be expressed as
  \begin{align}\nonumber
 \boldsymbol{\mathcal{Y}}_{\textrm{ss},l_1,l_2,l_3,l_4}&=\mathrm{fold}_{\langle 4 \rangle}[\boldsymbol{S}_{l_1,l_2,l_3,l_4}\boldsymbol{Y}_{\langle4 \rangle}]\\\nonumber
&=\boldsymbol{\mathcal{A}}\times_1 \boldsymbol{S}_{\hat{l}_1}\times_2 \boldsymbol{S}_{\hat{l}_2}\times_3 \boldsymbol{S}_{\hat{l}_3}\times_4 \boldsymbol{S}_{\hat{l}_4} \\
&=[\![\boldsymbol{\alpha};\boldsymbol{S}_{\hat{l}_1}\boldsymbol{U}_1, \boldsymbol{S}_{\hat{l}_2}\boldsymbol{U}_2, \boldsymbol{S}_{\hat{l}_3}\boldsymbol{U}_3, \boldsymbol{S}_{\hat{l}_4}\boldsymbol{U}_4]\!].
\end{align}
Concatenating $\boldsymbol{\mathcal{Y}}_{\textrm{ss},{l_1,l_2,l_3,l_4}}$ along the $5$-th dimension, we obtain a smoothed signal tensor, that is,
  \begin{align}\label{ss1}\nonumber
\boldsymbol{\mathcal{Y}}_{\textrm{ss}}=&[\boldsymbol{\mathcal{Y}}_{\textrm{ss},1,1,1,1}\sqcup_5...\sqcup_5\boldsymbol{\mathcal{Y}}_{\textrm{ss},L_1,1,1,1}\\\nonumber
&\boldsymbol{\mathcal{Y}}_{\textrm{ss},1,2,1,1}\sqcup_5...\sqcup_5\boldsymbol{\mathcal{Y}}_{\textrm{ss},L_1,2,1,1}...\\\nonumber
&\boldsymbol{\mathcal{Y}}_{\textrm{ss},1,L_2,L_3,L_4}\sqcup_5...\sqcup_5\boldsymbol{\mathcal{Y}}_{\textrm{ss},L_1,L_2,L_3,L_4}]\\
=&[\![\boldsymbol{\alpha};\boldsymbol{B}_1, \boldsymbol{B}_2, \boldsymbol{B}_3, \boldsymbol{B}_4, \boldsymbol{B}_5]\!].
\end{align}
According to \cite{sorensen2013blind}, it can be verified that $\boldsymbol{B}_n=\boldsymbol{S}_{\hat{l_n}}\boldsymbol{U}_n=\boldsymbol{U}_n^{J_n}\in \mathbb{C}^{J_n\times R}, n=1,2,3,4$, and $\boldsymbol{B}_5=\boldsymbol{U}_1^{L_1}\odot\boldsymbol{U}_2^{L_2}\odot\boldsymbol{U}_3^{L_3}\odot\boldsymbol{U}_4^{L_4}\in \mathbb{C}^{\prod_{k=1}^{4}L_k\times R}$.

\textbf{FBA:}
We also use FBA to further reduce the computational complexity.
As mentioned in \cite{485927}, the inner product between any two conjugate centro-symmetric vectors is real-valued, which permits to transform the complex-valued element space into a real-valued space.
In the case of 2D matrix, the FBA operation transforms the centro-Hermitian matrix $\boldsymbol{Z} \in \mathbb{C}^{M\times N}$ into $\boldsymbol{Z}_{\textrm{FB}} \in \mathbb{R}^{M\times 2N}$ by means of the unitary matrix $\boldsymbol{Q}_n \in \mathbb{C}^{n\times n}$, as follows:
\begin{align}
\boldsymbol{Z}_{\textrm{FB}}=\boldsymbol{Q}_M^{\mathrm{H}}[\boldsymbol{Z},~\boldsymbol{J}_M \boldsymbol{Z}^* \boldsymbol{J}_N]\boldsymbol{Q}_{2N} \in \mathbb{R}^{M\times 2N}.
\end{align}
Note that the matrix $\boldsymbol{Q}_{2n} \in \mathbb{C}^{2n\times 2n}$ may take on various forms.
Herein, the following one is used
\begin{align}
\boldsymbol{Q}_{2n} =\frac{1}{\sqrt{2}}  \begin{bmatrix}
 \boldsymbol{I}_n& j\boldsymbol{I}_n \\
 \boldsymbol{J}_n&  -j\boldsymbol{J}_n
\end{bmatrix},
\end{align}
and it satisfies the property that $\boldsymbol{J}_{2n}\boldsymbol{Q}_{2n}=\boldsymbol{Q}_{2n}^*$.
As an analogy to the matrix case, the complex tensor-based FBA can be deduced by first transforming $\boldsymbol{\mathcal{Y}}_{\textrm{ss}}$ into a centro-Hermitian tensor $\boldsymbol{\mathcal{Y}}_{\textrm{CH}}$, that is,
\begin{align}\nonumber
\boldsymbol{\mathcal{Y}}_{\textrm{CH}}=&\left[\boldsymbol{\mathcal{Y}}_{ss} \sqcup_5 \left[\boldsymbol{\mathcal{Y}}_{ss}^*\times_1 \boldsymbol{J}_{J_1} \times_2 \boldsymbol{J}_{J_2}...\times_5 \boldsymbol{J}_{J_5} \right] \right]\\
 &\in \mathbb{C}^{J_1\times J_2\times J_3\times J_4 \times 2J_5 },
\end{align}
where $J_5=\prod_{k=1}^{4}L_k$.
Next, this centro-Hermitian tensor is mapped to a real-valued tensor $\boldsymbol{\mathcal{Y}}_{\textrm{FB}}$, that is,
\begin{align}\nonumber
\boldsymbol{\mathcal{Y}}_{\textrm{FB}}=&\boldsymbol{\mathcal{Y}}_{\textrm{CH}}\times_1 \boldsymbol{Q}_{J_1}^{\mathrm{H}} \times_2 \boldsymbol{Q}_{J_2}^{\mathrm{H}}\times_3 \boldsymbol{Q}_{J_3}^{\mathrm{H}}\times_4 \boldsymbol{Q}_{J_4}^{\mathrm{H}}\times_5 \boldsymbol{Q}_{J_5}^{\mathrm{T}}\\
 &\in \mathbb{C}^{J_1\times J_2\times J_3\times J_4 \times 2J_5 }.
\end{align}

\subsection{Parameter Estimation via Rotational Invariance Property}\label{sec3.3}
Based on the preprocessing step, we then reconstruct the signal tensor in (\ref{ss1}), which allows us to use the ESPRIT algorithm.
A key step of ESPRIT is to exploit the rotational invariance property \cite{485927}.

Based on the real-valued tensor, we now extend the classical matrix version of unitary ESPRIT to tensor version.
Using the relationship between the CPD and TT decomposition \cite{10048567}, and the extension from matrix to tensor \cite{4545266}, one can deduce that the vector spaces spanned by the $r$-th mode vectors of $\boldsymbol{\mathcal{G}}$ and $\boldsymbol{\mathcal{F}}$ are the same for $r=1,2,3,4$, where $\boldsymbol{\mathcal{G}}=\boldsymbol{\mathcal{G}}_1\times^1_2 \boldsymbol{\mathcal{G}}_2 \times^1_3 \boldsymbol{\mathcal{G}}_3 \times^1_4 \boldsymbol{\mathcal{G}}_4 \in \mathbb{C}^{I_1\times I_2\times I_3\times I_4\times R }$ and $\boldsymbol{\mathcal{F}}=(\boldsymbol{Q}_{J_1}^{\mathrm{H}}\boldsymbol{B}_1)\odot(\boldsymbol{Q}_{J_2}^{\mathrm{H}}\boldsymbol{B}_2)\odot(\boldsymbol{Q}_{J_3}^{\mathrm{H}}\boldsymbol{B}_3)\odot(\boldsymbol{Q}_{J_4}^{\mathrm{H}}\boldsymbol{B}_4)\in \mathbb{C}^{J_1\times J_2\times J_3\times J_4\times R }$.
This property enables us to establish the real-valued rotational invariance equations.

Specifically, given two selection matrices $\boldsymbol{S}^{(1)}_n\triangleq[\boldsymbol{I}_{I_n-1}, \boldsymbol{0}_{I_n-1,1}]$, $\boldsymbol{S}^{(2)}_n\triangleq[\boldsymbol{0}_{I_n-1,1},\boldsymbol{I}_{I_n-1}]$, one can obtain that
\begin{align}
\boldsymbol{\mathcal{G}}\times_n\boldsymbol{K}^{(1)}_n\times_5 \boldsymbol{\Upsilon}_n\approx\boldsymbol{\mathcal{G}}\times_n\boldsymbol{K}^{(2)}_n, n=1,2,3,4,
\end{align}
where $\boldsymbol{K}^{(1)}_n\triangleq\mathrm{Re}\{\boldsymbol{Q}_{I_n-1}^{\mathrm{H}}  \boldsymbol{S}^{(1)}_n \boldsymbol{Q}_{I_n}\}$ and $\boldsymbol{K}^{(2)}_n\triangleq\mathrm{Im}\{\boldsymbol{Q}_{I_n-1}^{\mathrm{H}}  \boldsymbol{S}^{(1)}_n \boldsymbol{Q}_{I_n}\}$.
Similar to the matrix approach, $\boldsymbol{\Upsilon}_n\in \mathbb{C}^{R\times R},~  n=1,2,3,4$ are related to the eigenvalue transformation, which are the asymptotic estimates of $\tan(\Theta_r/2)$, $\tan(\Phi_r/2)$, $\tan(\eta_r/2)$, and $\tan(\mu_r/2)$.
 The solution for $\boldsymbol{\Upsilon}_n$ is given as
 \begin{align}
\boldsymbol{\Upsilon}_n=\left\{\left[\boldsymbol{\hat{K}}^{(1)}_n \boldsymbol{G}^{\mathrm{T}}_{(5)}\right]^{\dagger}\boldsymbol{\hat{K}}^{(2)}_n  \boldsymbol{G}^{\mathrm{T}}_{(5)}\right\}^{\mathrm{T}},\label{paraest}
\end{align}
where $\boldsymbol{\hat{K}}^{(1)}_n$ and  $\boldsymbol{\hat{K}}^{(2)}_n$ are extensions of $\boldsymbol{{K}}^{(1)}_n$ and  $\boldsymbol{{K}}^{(2)}_n$, that is,
$\boldsymbol{\hat{K}}^{(1)}_n=\boldsymbol{I}_{\prod_{k=1}^{n-1}I_k}\otimes\boldsymbol{K}^{(1)}_n\otimes \boldsymbol{I}_{\prod_{k=n+1}^{4}I_k}$,
$\boldsymbol{\hat{K}}^{(2)}_n =\boldsymbol{I}_{\prod_{k=1}^{n-1}I_k}\otimes\boldsymbol{K}^{(2)}_n \otimes\boldsymbol{I}_{\prod_{k=n+1}^{4}I_k}$.

\subsection{Estimation of the Target Parameters}\label{sec3.4}
To finally estimate the target parameters, the only calculation needed is to obtain the eigenvalues of $\boldsymbol{\Upsilon}_n$.
Many methods can be used to calculate the eigenvalues of $\boldsymbol{\Upsilon}_n$ such as the simultaneous Schur decomposition (SSD) \cite{haardt1998simultaneous}, eigenvalue decomposition, etc.
However, the correct pairing of the parameters over different modes should be guaranteed.
Thus, SSD is used to estimate the joint eigenvalues and, correspondingly, the spatial frequencies of the $r$-th target.
Denote $\lambda_n(r)$ as the $r$-th eigenvalue of $\boldsymbol{\Upsilon}_n$, then we have $\lambda_1(r)=\tan(\Theta_r/2)$, $\lambda_2(r)=\tan(\Phi_r/2)$, $\lambda_3(r)=\tan(\eta_r/2)$, $\lambda_4(r)=\tan(\mu_r/2)$.
Combined with (\ref{a1})-(\ref{a4}), the distance, the velocity, the azimuth and the elevation directions of departure for the $r$-th target are estimated respectively as
\begin{align}\nonumber
\hat{R}_r=\frac{\pi\eta_rc}{\Delta f T_s},&~\hat{v}_r=\frac{2\pi\mu_r \lambda}{T},\\
\hat{\phi}_r=\arctan\left(\frac{\Phi_r}{\Theta_r}\right),&~\hat{\theta}_r=\arcsin\left(\frac{\lambda}{d}\sqrt{\Theta_r^2+\Phi_r^2} \right). \label{estimatedPara}
\end{align}
\subsection{Computational Complexity}\label{sec3.5}
The computational complexity of the joint parameter estimation algorithm comes from data smoothing and the use of rotational invariance property.
For data smoothing, the computational complexities of SS and FBA are respectively $\mathcal{O}(\sum_{n=1}^{4}I_nT_nR)$ and $\mathcal{O}(\Pi_{n=1}^{5}J_n\cdot\sum_{n=1}^{5}J_n)$, where the complexity of generating index is ignored.
Similar to the classical matrix version of ESPRIT, the computational complexity associated with using the rotational invariance property mainly comes from calculating (\ref{paraest}), which is $\mathcal{O}(\Pi_{n=1}^{4}I^2_nR+\Pi_{n=1}^{4}I_nR^2+R^3)$
   by regarding $\mathcal{O}((I_n-1)/I_n\Pi_{n=1}^{4}I_n)$ and $\mathcal{O}(\Pi_{n=1}^{4}I_n)$ being the same.
Therefore, the total computational complexity of the parameter estimation is $\mathcal{O}(\Pi_{n=1}^{5}J_n\cdot\sum_{n=1}^{5}J_n+\Pi_{n=1}^{4}I^2_nR+\Pi_{n=1}^{4}I_nR^2)$, where the items of power below 6 are ignored.

\section{Numerical and Experimental Evaluations}\label{Result}
In this section, we present results of exemplary simulation as well as experimental results based on real measurements to evaluate the effectiveness of the proposed method.
We first describe the radar system configurations for the simulation and field measurement collection.
Then, the effectiveness of the proposed denoising scheme is demonstrated in terms of SNR evaluations as well as visual evaluations.
Better denoising leads to improved target estimation achieved by exploiting the information about higher-order signal tensor structure.
Then, the ESPRIT-based method for parameter estimation is implemented to estimate the target parameters simultaneously.
\vspace{-1em}
\begin{table*}[!]
\centering
\caption{{System Parameters}}\label{tab:PlatformPara}
\begin{tabular}{ll|ll|ll}
\cline{1-6}
\toprule[2pt]
  Parameter (for simulation)                & Value &  Parameter (for measurement)                & Value &Parameter (for both)                    & Value               \\ \cline{1-6}
\specialrule{0em}{1pt}{1.5pt}
  Tx elements ($I_1$)       & 9    &  Tx elements ($I_1$)       & 2& Center Frequency     & 77 GHz               \\ \cline{1-6}
\specialrule{0em}{1pt}{1.5pt}
  Rx elements ($I_2$)    & 25 &  Rx elements ($I_2$)    & 4 &Chirp Slope     & 85.17 MHz/$\mu$s      \\      \cline{1-6}
\specialrule{0em}{1pt}{1.5pt}
    {Samples/ Chirp } ($I_3$)       & 256  &  {Samples/ Chirp } ($I_3$)       & 186        &Sampling Rate                &   6300 ksps      \\\cline{1-6}
 \specialrule{0em}{1pt}{1.5pt}
   {Chirp/ Frame} ($I_4$)     & 128     & {Chirp/ Frame} ($I_4$)     & 128    &Bandwidth             & 2.51 GHz        \\
  \specialrule{0em}{1.5pt}{1.5pt}
\cline{1-6}
\end{tabular}\vspace{-1em}
\end{table*}
\begin{figure}[!t]
\centering
\vspace{-0em}
  \includegraphics[width=0.75\columnwidth,draft=false]{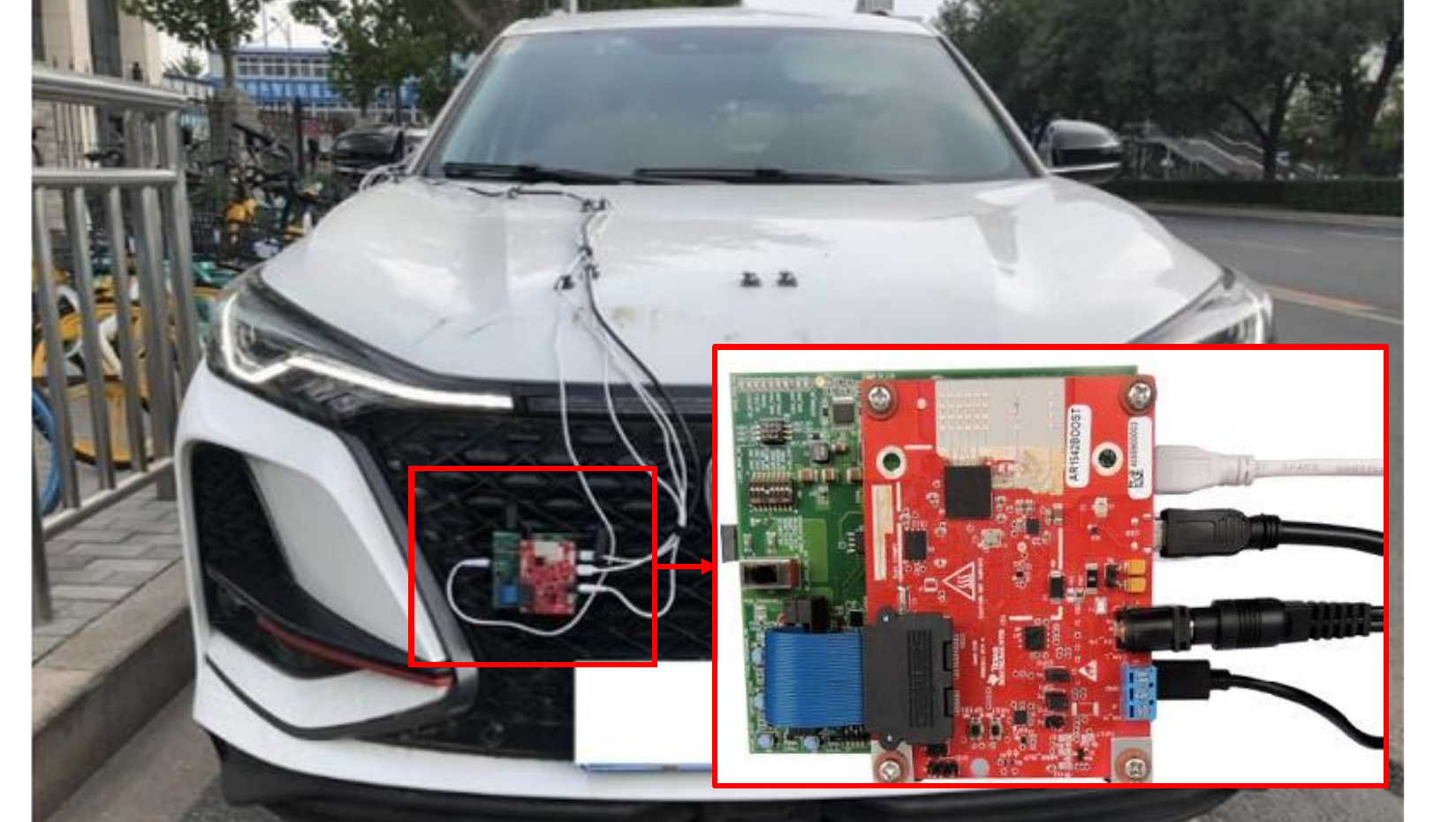}
\caption{\small Measurement setups: radar board hardware and test SUV.}\label{TestingPlatformfig}\vspace{-1em}
\end{figure}
\subsection{Configurations of Radar System and Measurement}
To show the performance of the proposed method in practical systems, field measurements with an optimized radar experimental setup were carried out in Haidian District, Beijing, to capture data during parallel parking for experimental evaluation.
 Table \ref{tab:PlatformPara} also shows the radar system parameters used while collecting measurements.
 Note that the sample numbers under per chirp duration for real measurement is different from that used for simulations, while the number of points for computing FFT along the dimension is set to be 256 by zero padding operation applied for both.
 Moreover, the antenna arrays at the transmit and receive sides for collecting measurements are $2\times 1$ ($K_{T_a}= 2, K_{T_e}=1$) and $4\times1$ ($K_{R_a}=4, K_{R_e}=1$), respectively, yielding an $8\times1$ virtual linear array with $\lambda/2$ spacing between elements.
 The test SUV, coming with the radar board hardware, a dashcam and global positioning system (GPS), records the situation and the realtime position of the vehicle as ground truth for the subsequent data processing.
The radar board hardware shown in red square of Fig. \ref{TestingPlatformfig} adopts the AWR1843 evaluation module and DCA1000EVM real-time data capture card developed by Texas Instruments.
Note that another existing hardware in the lab is IWR 6843 evaluation module, which is available for 4D data capturing.
However, it necessitates decomposing the angular processing into two separate 1D problems with elevation and azimuth separately, as the three transmitters display the shape of ``L" while the four receivers display a $2\times 2$ array.
 Such configuration leads to inherently two 3D signal processing (range-Doppler-azimuth/elevation angle) under TDM, where the algorithm works with two distinct 3D datasets rather than a unified 4D tensor structure.
The limitation makes IWR 6843 not optimal to exploit the full tensor structure of the data compared to AWR 1843, as the working frequency of IWR 6843 is 60-64 GHz below the automotive frequency band.
While the proposed algorithm is designed for 4D signal processing, our experimental results thus demonstrate the algorithm's performance under current hardware constraints.
Nevertheless, we further show that the enhanced capability of the proposed algorithm for 4D data would remove noise components and improve estimation accuracy by preserving the complete multilinear relationships.
These findings suggest that there is a strong incentive to add the corresponding flexibility to future hardware advancements, which should prioritize native 4D sensing capabilities to fully unlock the potential of tensor-based signal processing algorithms.
However, the development of new hardware falls beyond the scope of this work and typical university research capabilities.
\begin{figure}[!t]
\centering
\includegraphics[width=0.8\columnwidth,draft=false]{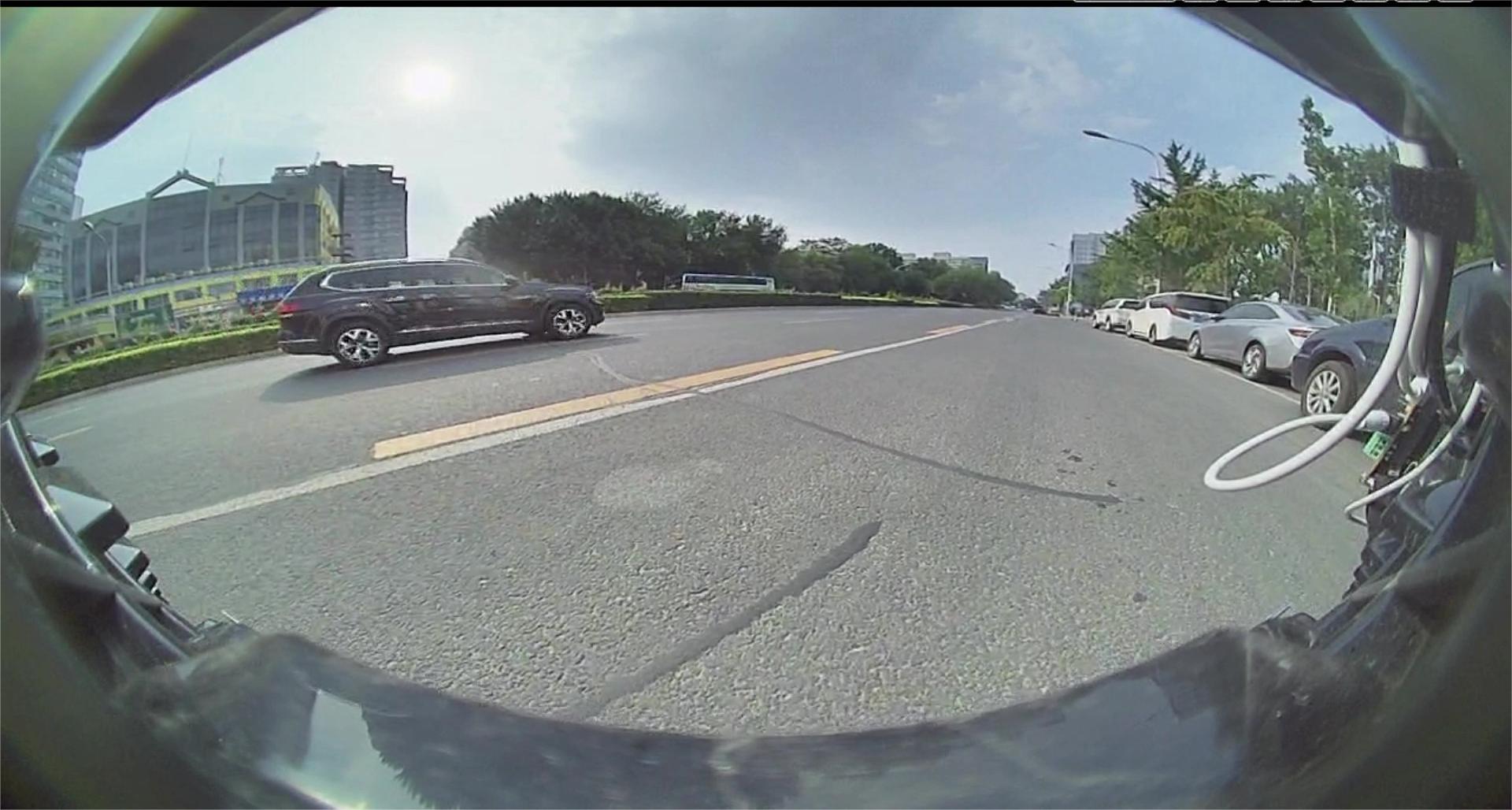}
\caption{\small Sample image of the selected frame.\label{Frame87TestCut}}\vspace{-2em}
\end{figure}

\begin{figure*}[!t]
\vspace{-0em}
\center
\includegraphics[width=1.7\columnwidth,draft=false]{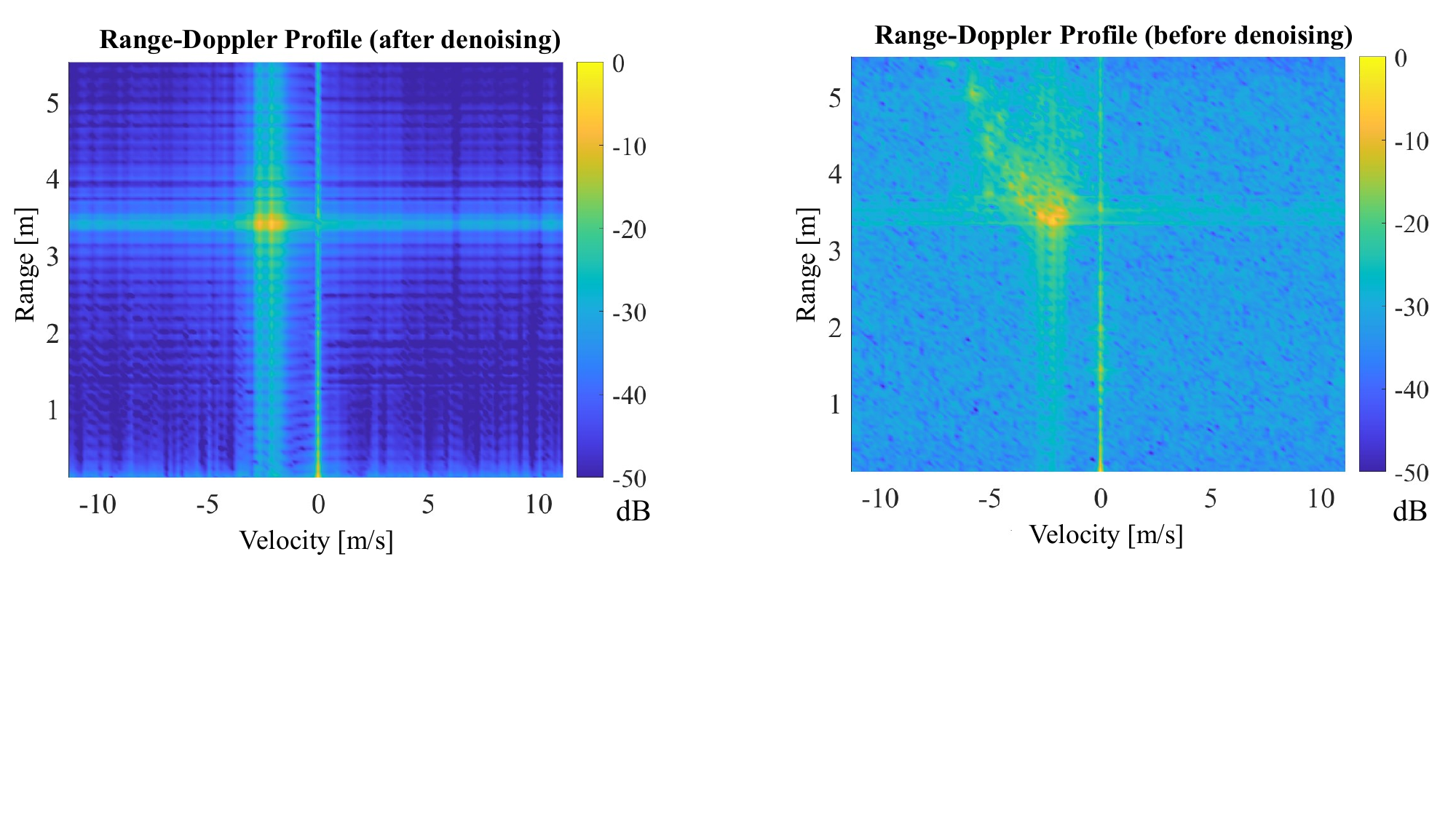}\\\vspace{-8em}
\caption{\small Denoising performance comparisons of range-Doppler profiles. The value of colorbar is normalized to the maximum value. \label{RDMeas}}\vspace{-1em}
\end{figure*}

\begin{figure*}[!t]
\vspace{-0em}
\center
\includegraphics[width=1.7\columnwidth,draft=false]{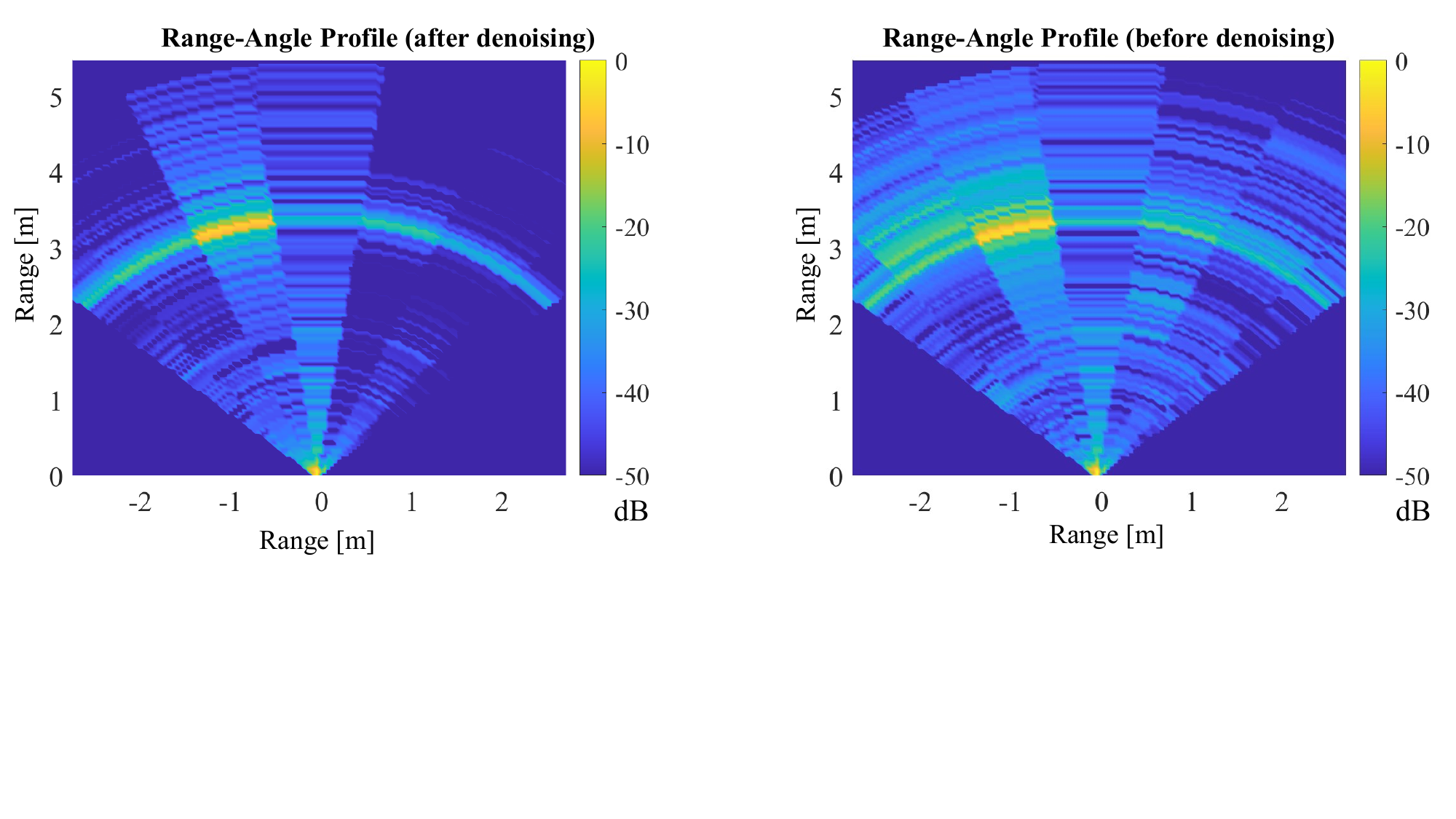}\\\vspace{-8em}
\caption{\small Denoising performance comparisons of range-angle (azimuth) profiles. The value of colorbar is normalized to the maximum value. \label{RAMeas}}\vspace{-1em}
\end{figure*}
\begin{figure*}[!t]
\center
\includegraphics[width=1.9\columnwidth,draft=false]{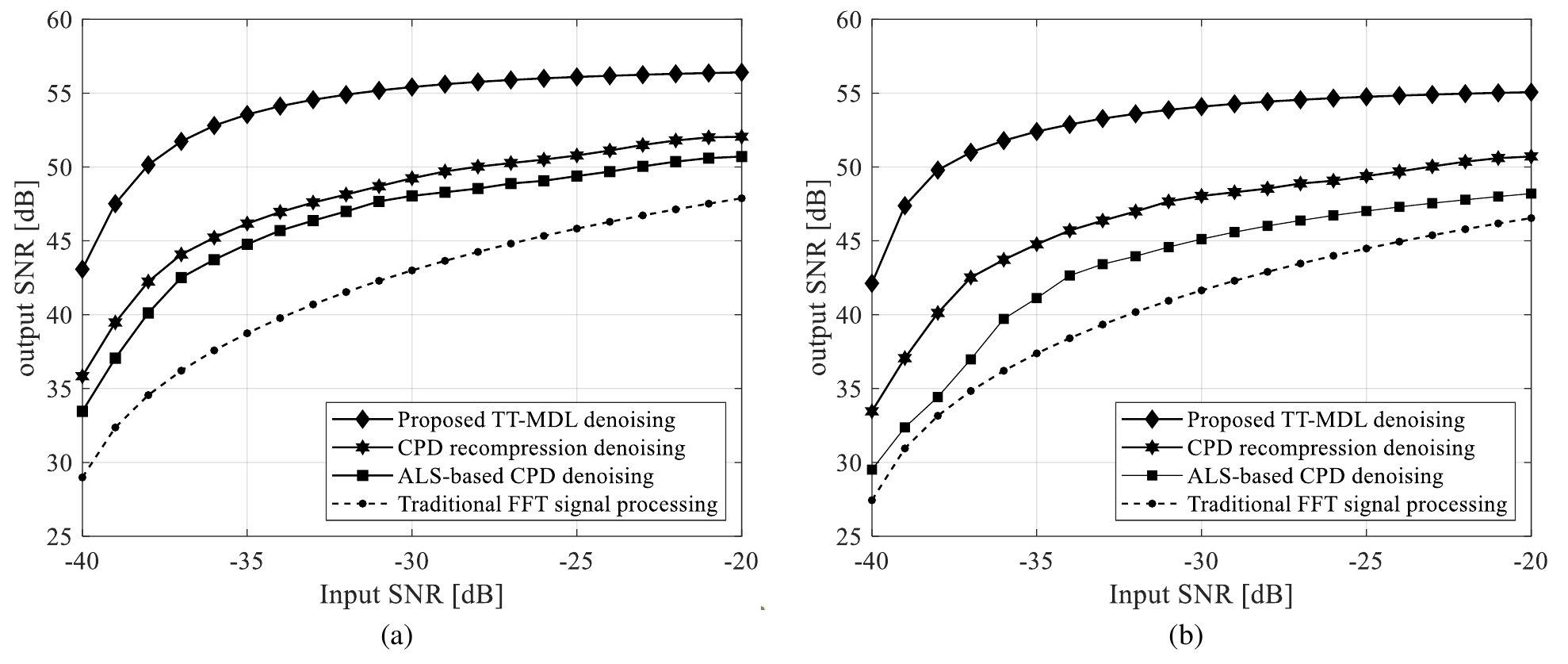}\\
\caption{\small Output SNR vs. input SNR.
The received power of both targets is enhanced by different denoising methods for two targets with the same range as a coherent parameter, while the Doppler and spatial locations of the targets are different.
(a) The case with a relatively strong received power.
(b) The case with a relatively weak received power.
\label{SNRforT1T2}}\vspace{-2em}
\end{figure*}
 To better demonstrate the effectiveness of the 4D method, computer simulations with the radar system configurations shown in Table~\ref{tab:PlatformPara} are also performed.
  We consider an FMCW automotive radar system with TDM-MIMO configuration, where the URAs at the transmit and receive sides are $3\times3$ ($K_{T_a}= K_{T_e}=3$) and $5\times5$ ($K_{R_a}=K_{R_e}=5$), respectively, yielding a $15\times15$ virtual URA with $\lambda/2$ spacing between elements.
  The signal reflected by the scatter point of the target is sampled in both fast and slow time dimensions and aggregated in each virtual element to generate the received signal tensor $\boldsymbol{\mathcal{Y}}$.
  Herein, the analysis focuses on the target's appearance at the range-Doppler-azimuth-elevation angle of {(24 m, 12m/s, 17.5 deg, 56.3 deg) and (24 m, -13m/s, -36.8 deg, 36.9 deg)}.
    The above results are used to illustrate the case where the coherent signals are backscattered from two targets that have two different angles, but which are located on the same distance to the radar system.
  The above results are used to illustrate the case involving two targets located at the same distance from the radar system, which results in coherent signals backscattered from different angles.
    The performance of the proposed method is evaluated by averaging 1000 independent Monte Carlo trials.
 The above settings are used in our experiments unless otherwise specified.

 To evaluate the effectiveness of noise reduction and parameter estimation, two performance metrics are used:
the output SNR of the target and the joint normalized mean square error (NMSE) of the estimated parameters.

\subsection{Denoising Results}
 \begin{figure}[t]
\center
\includegraphics[width=0.8\columnwidth,draft=false]{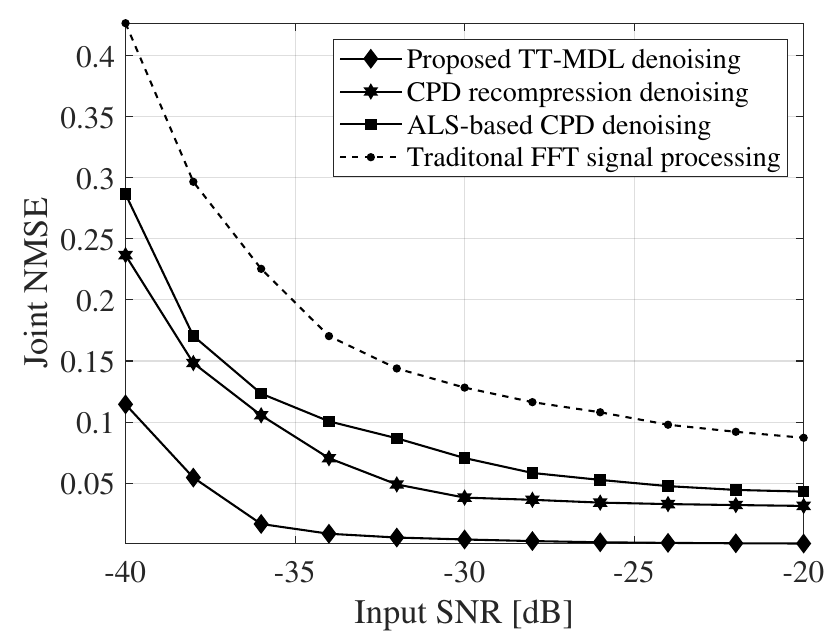}\\
\caption{\small Joint NMSE for estimated parameters of the stronger target under different input SNR conditions. \label{JointNMSE}}\vspace{-2em}
\end{figure}
For the field measurements, we apply the denoising and parameter estimation methods.
Since real targets are extended and non-point-shaped, it is more likely that more than one scattering point, i.e., sampled signal elements, correspond to each target, which leads to a higher estimated TT rank compared to simulation example.
In Fig. \ref{Frame87TestCut}, the black SUV corresponds to the target shown in Figs. \ref{RDMeas} and \ref{RAMeas}.
The target, in practice, is often appearing as more than one scattering point, so that the TT rank calculated by (\ref{MDL}) in this case is three.
By visual inspection of Figs. \ref{RDMeas} and \ref{RAMeas}, it is clear that the noise reduction using the proposed denoising algorithm leads to better target recognition.
 It also shows lower noise level from the background.
The denoising process removes the unwanted components from the received signal, hence, facilitating the estimation of target parameters.

The numerical results in Fig. \ref{SNRforT1T2} demonstrate the effectiveness of the proposed TT-MDL denoising method compared to CPD recompression, ALS-based CPD, and traditional FFT signal processing methods for both cases of strong and weak received power.
Two targets at identical ranges are set in the simulation, where the reflected signals from the targets would be coherent.
By exploiting the TT decomposition's hierarchical rank adaptivity, the proposed method achieves consistent SNR enhancement across both strong and weak target cases as
can be seen respectively in Figs. \ref{SNRforT1T2}(a) and \ref{SNRforT1T2}(b).
  The output SNR improvements are 9-16 dB compared to traditional FFT processed signals that manifest SNR gains from coherent superposition from Fourier operations.
In addition, the TT-MDL framework leverages MDL-based adaptive rank selection and TT decomposition to preserve coherent parameters (e.g., shared target range) while suppressing noise, which explains its superior performance in structurally complex scenarios in contrast to CPD recompression and ALS-based CPD methods.
For instance, in the strong-power case in Fig. \ref{SNRforT1T2}(a), TT-MDL attains about 56~dB at an input SNR of -20 dB, outperforming ALS-based CPD (about 52~dB) by 4 dB and CPD recompression (about 50~dB) by 6~dB.
This advantage persists in the weak-power case in Fig.~ \ref{SNRforT1T2}(b), where TT-MDL yields about 55~dB at the same input SNR, surpassing ALS-based CPD (about 50~dB) by 5~dB and CPD recompression (about 48~dB) by 7~dB.

Notably, the TT-MDL framework addresses the critical limitations of ALS-based CPD.
Firstly, CPD's global CP-rank inherently preserves noise components in coherent dimensions, whereas QR-enabled recompression based on TT-format selectively truncates these noise subspaces through localized singular value thresholding.
This can be deduced by comparing the gaps between ALS-based CPD and traditional FFT processed curves for two targets, especially when the input SNRs are low.
Secondly, ALS-based CPD exhibits degraded performance for weak targets in Fig. \ref{SNRforT1T2}(b), while TT-MDL maintains robustness due to its MDL-driven rank adaptation, which autonomously balances low-rank signal extraction against residual noise levels without requiring prior noise estimates.
The observed output SNR gains at higher input SNRs beyond -30 dB further validates the method's adaptability.
 As inherent noise power decreases the residual noise becomes negligible compared to the signal power, leaving limited room for further SNR improvement, which can be observed in all curves.
These results validate TT-MDL as a robust solution for multi-target denoising, balancing quantitative SNR gains with stable performance.

\begin{table}[!t]
\centering
\caption{Time Consumption of the Algorithm}\vspace{-0em}
\begin{tabular}{c|c|c|c}\cline{1-4}
\toprule[2pt]
Procedure & TT-MDL  & Data smoothing & Parameter estimation \\\cline{1-4}\specialrule{0em}{1pt}{1.5pt}
Run Time (ms)& 26.9  & 21.3         & 18.6 \\
\bottomrule[2pt]\cline{1-4}
\end{tabular}\vspace{-2em}\label{time}
\end{table}

\subsection{Performance of Parameter Estimation}
The denoising removes the unwanted components from the received signal, hence, facilitating the estimation of target parameters.
The target parameters are jointly estimated by the proposed ESPRIT-based method with the range-velocity-angle approximately of (3.4 m, -2.5 m/s, -14 deg), which agrees with the visual results in Figs. \ref{RDMeas} and \ref{RAMeas}.
It can be seen that the proposed ESPRIT-based method shows excellent performance in estimating parameters simultaneously.
Table \ref{time} demonstrates the rum time of each procedure in the algorithm to show the potential for real-time processing.
The processor type is 11th Gen Intel(R) Core(TM) i5-1145G7 and the installed RAM is 16 GB.
Thus, it can be implemented in real automotive radar systems to provide high accuracy estimation of location and velocity of targets in real time.

To illustrate the parameter estimation performance more distinctly, the performance of joint parameter estimation using the proposed $4$-order tensor is evaluated for the denoised and traditional FFT processed signals versus input SNR.
In addition, the joint NMSE across the parameters of the $r$-th target, namely, distance, velocity, azimuth angle, and elevation angle, is utilized in the Monte Carlo simulations to assess the estimation performance.
The NMSE is given by
\begin{align}\nonumber
\mathrm{NMSE}_r=&\left(\frac{\hat{R}_r-R_r}{R_r}\right)^2+\left(\frac{\hat{v}_r-v_r}{v_r}\right)^2\\
&+\left(\frac{\hat{\phi}_r-\phi_r}{\phi_r}\right)^2+\left(\frac{\hat{\theta}_r-\theta_r}{\theta_r}\right)^2,
\end{align}
where $\hat{R}_r$, $\hat{v}_r$, $\hat{\phi}_r$, and $\hat{\theta}_r$ are the estimated parameters obtained by (\ref{estimatedPara}), and ${R}_r$, ${v}_r$, ${\phi}_r$, and ${\theta}_r$ are the true parameters of the target.
As shown in Fig. \ref{JointNMSE}, the proposed TT-MDL denoising framework achieves superior joint NMSE performance for estimating parameters across input SNR levels.

One can observe that the ESPRIT-based estimation accuracy improves as the SNR increases for both targets, which shows that the estimates of the parameters are asymptotically unbiased.
The improved estimation accuracy can be attributed to the effective noise suppression and structural preservation enabled by the TT-MDL denoising framework.
Unlike CPD-based methods, which suffer from fixed-rank approximation errors and noise-sensitive ALS convergence, the proposed approach maintains the multilinear signal structure critical for accurate parameter estimation.
For instance, TT-MDL attains the lowest NMSE values outperforming other frameworks.
Specifically, in the case of -40 dB input SNR, the joint NMSE under the TT-MDL framework is approximately 0.11, significantly lower than CPD recompression (0.24), ALS-based CPD (0.28), and the FFT processed baseline (0.43).
Notably, even under extreme noise of -40 dB input SNR, TT-MDL reduces NMSE by 75\% compared to FFT processed framework, demonstrating robust noise-resistance without sacrificing signal fidelity.
The above results highlight the robustness of the parameter estimation algorithm in the proposed framework, which leverages the denoised signal structure to achieve reliable performance even under extreme noise conditions.
In addition, the capability of jointly pairing and estimating the parameters makes it more efficient for practical utilizations where accurate parameter estimation is paramount.
\section{Conclusion}\label{conclusion}
To enhance target localization for FMCW MIMO radar system in noise-polluted environments, we propose a joint noise reduction and parameter estimation framework based on tensor decomposition.
By employing TT decomposition with adaptive rank selection, we effectively overcome the global rank constraints of conventional CPD methods while achieving low-rank extraction without noise-dependent prior.
Then, we perform the integrated spatial smoothing and FBA pre-processing to mitigate rank deficiency issues for coherent signals, and exploit the tensor-ESPRIT extension to enable accurate joint 4D parameter estimation simultaneously without computationally demanding eigen-decomposition or peak searching.
Simulation results have illustrated that the proposed framework gives output SNR improvements of 9-16~dB compared to traditional FFT signal processing in the scenario with two coherent targets, and provides accurate parameter estimation performance with 75\% NMSE reduction under TT-MDL denoising compared to FFT processed framework.
Practical results have approved that our method is suitable for automotive applications, facilitating the deployment of MIMO FMCW systems in practical scenarios.

\section*{Appendix A}

The equivalence between the core tensor $\boldsymbol{\mathcal{A}} \in \mathbb{C}^{R \times R...\times R}$ in CPD and the TT cores is established in \emph{{Theorem 1}}.
Similar to (\ref{TT}) and (\ref{TTDele}), the core tensor can by rewritten as
\begin{align}
\boldsymbol{\mathcal{A}}
=\boldsymbol{I}_{T_1}\times^1_2\boldsymbol{\mathcal{T}}_{2}\times^1_3...\times^1_{\bar{n}}\boldsymbol{\mathcal{T}}_{\bar{n}}\times^1_{\bar{n}+1}...\times^1_{N-1}\bar{\boldsymbol{\mathcal{T}}}_{N-1}\times^1_N\boldsymbol{I}_N.
\end{align}
Then, following the definition of the CPD in (\ref{CPD}), we obtain the CPD in Subsection \ref{CPD2TT} by replacing the above TT decomposition of $\boldsymbol{\mathcal{A}}$ in \emph{Theorem} 1. $\hfill\blacksquare$
\section*{Appendix B}
Recall that the CPD mode-$n$ unfolding matrix, which is of size $\prod_{l=1}^{n}I_l\times \prod_{l={n+1}}^{N}I_l$, is given by
\begin{equation}
{\boldsymbol{X}}_{\langle n \rangle}=(\boldsymbol{U}_{n}\odot...\odot\boldsymbol{U}_{1})\mathrm{diag}(\mathbf{\alpha}) (\boldsymbol{U}_N\odot...\odot\boldsymbol{U}_{n+1})^{\mathrm{T}}.
\end{equation}
Based on the matricization form and exploiting the Vandermonde structure, it is straightforward to deduce that the complex generators for $\boldsymbol{U}_{n}$, $n=1,2,3,4$, are drawn from the parameter space $\mathbb{C}^{R}$.
According to \cite{jiang2001almost}, if the parameter distribution is continuous with respect to the Lebesgue measure in $\mathbb{C}^{R}$, the Khatri-Rao product of all $\boldsymbol{U}_n$ appears to exhibit full rank almost surely.
Equipped with the above result, we can obtain $\mathrm{rank}(\boldsymbol{X}_{\langle n \rangle})=\min(\prod_{l=0}^{n-1}I_l,R,\prod_{l=n}^{N}I_l)$, $n=1, 2,..., N$.
Then, following \emph{Theorem} 2 in Section 3 of \cite{oseledets2011tensor}, we can directly deduce that the rank $T_n$ satisfies $T_n\geq\mathrm{rank}(\mathrm{X}_{\langle n \rangle})$, $n=1, 2,..., N$, which is known as a \emph{rank deficient} case.
In other words, TT decomposition shows a \emph{rank deficient} structure that allows to exploit the low rank property of $\boldsymbol{\mathcal{X}}_{\mathrm{CP}}$, which is captured in the CP tensor model, even if the tensor $\boldsymbol{\mathcal{X}}_{\mathrm{CP}}$ under TT decomposition has a full rank.
$\hfill\blacksquare$

\bibliographystyle{IEEEtran} \small
\bibliography{IEEEabrv,Globecom}

\begin{thebibliography}{10}
\providecommand{\url}[1]{#1}
\csname url@samestyle\endcsname
\providecommand{\newblock}{\relax}
\providecommand{\bibinfo}[2]{#2}
\providecommand{\BIBentrySTDinterwordspacing}{\spaceskip=0pt\relax}
\providecommand{\BIBentryALTinterwordstretchfactor}{4}
\providecommand{\BIBentryALTinterwordspacing}{\spaceskip=\fontdimen2\font plus
\BIBentryALTinterwordstretchfactor\fontdimen3\font minus
  \fontdimen4\font\relax}
\providecommand{\BIBforeignlanguage}[2]{{%
\expandafter\ifx\csname l@#1\endcsname\relax
\typeout{** WARNING: IEEEtran.bst: No hyphenation pattern has been}%
\typeout{** loaded for the language `#1'. Using the pattern for}%
\typeout{** the default language instead.}%
\else
\language=\csname l@#1\endcsname
\fi
#2}}
\providecommand{\BIBdecl}{\relax}
\BIBdecl

\bibitem{he2023physics}
D.~He, K.~Guan, D.~Yan, H.~Yi, Z.~Zhang, X.~Wang, Z.~Zhong, and N.~Zorba,
  ``Physics and {AI-based} digital twin of multi-spectrum propagation
  characteristics for communication and sensing in {6G} and beyond,''
  \emph{IEEE J. Sel. Areas in Communications}, vol.~41, no.~11, pp. 3461--3473,
  2023.

\bibitem{10457955}
S.~Ren, Z.~Lei, Z.~Wang, M.~Dianati, Y.~Wang, S.~Chen, and W.~Zhang,
  ``Interruption-aware cooperative perception for {V2X} communication-aided
  autonomous driving,'' \emph{IEEE Trans. Intelligent Vehicles}, vol.~9, no.~4,
  pp. 4698--4714, 2024.

\bibitem{20214D}
S.~Sun and Y.~D. Zhang, ``4{D} automotive radar sensing for autonomous
  vehicles: A sparsity-oriented approach,'' \emph{IEEE J. Sel. Topics in Signal
  Processing}, vol.~15, no.~4, pp. 879--891, 2021.

\bibitem{10044244}
Y.~Wang, Q.~Zhang, Z.~Wei, L.~Kui, F.~Liu, and Z.~Feng, ``Performance analysis
  of uncoordinated interference mitigation for automotive radar,'' \emph{IEEE
  Trans. Vehicular Technology}, vol.~72, no.~4, pp. 4222--4235, 2023.

\bibitem{9760104}
A.~Venon, Y.~Dupuis, P.~Vasseur, and P.~Merriaux, ``Millimeter wave {FMCW
  RADAR}s for perception, recognition and localization in automotive
  applications: A survey,'' \emph{IEEE Trans. Intelligent Vehicles}, vol.~7,
  no.~3, pp. 533--555, 2022.

\bibitem{9853513}
Z.~Hu, J.~Huang, D.~Hu, and Z.~Wang, ``A time-frequency image denoising method
  via neural networks for radar waveform recognition,'' \emph{IEEE
  Communications Lett.}, vol.~27, no.~1, pp. 150--154, 2023.

\bibitem{9978673}
T.~Hao, L.~Jing, and W.~He, ``An automated {GPR} signal denoising scheme based
  on mode decomposition and principal component analysis,'' \emph{IEEE
  Geoscience and Remote Sensing Lett.}, vol.~20, pp. 1--5, 2023.

\bibitem{9521673}
S.~Tan, X.~Zhang, H.~Wang, L.~Yu, Y.~Du, J.~Yin, and B.~Wu, ``A {CNN}-based
  self-supervised synthetic aperture radar image denoising approach,''
  \emph{IEEE Trans. on Geoscience and Remote Sensing}, vol.~60, pp. 1--15,
  2022.

\bibitem{10164024}
L.~Zhu, Y.~Liu, D.~He, K.~Guan, J.~Liao, and Z.~Zhong, ``A low-complexity noise
  reduction algorithm for enhanced target detection in {FMCW} radar,''
  \emph{IEEE Trans. Vehicular Technology}, vol.~72, no.~12, pp.
  15\,227--15\,236, 2023.

\bibitem{xu2021transmit}
F.~Xu, S.~A. Vorobyov, and F.~Yang, ``Transmit beamspace {DDMA} based
  automotive {MIMO} radar,'' \emph{IEEE Trans. Vehicular Technology}, vol.~71,
  no.~2, pp. 1669--1684, 2021.

\bibitem{9495264}
H.~Zheng, Z.~Shi, C.~Zhou, M.~Haardt, and J.~Chen, ``Coupled coarray tensor
  {CPD} for {DOA} estimation with coprime {L}-shaped array,'' \emph{IEEE Signal
  Processing Lett.}, vol.~28, pp. 1545--1549, 2021.

\bibitem{1143830}
R.~Schmidt, ``Multiple emitter location and signal parameter estimation,''
  \emph{IEEE Trans. on Antennas and Propagation}, vol.~34, no.~3, pp. 276--280,
  1986.

\bibitem{32276}
R.~Roy and T.~Kailath, ``{ESPRIT}-estimation of signal parameters via
  rotational invariance techniques,'' \emph{IEEE Trans. on Acoustics, Speech,
  and Signal Processing}, vol.~37, no.~7, pp. 984--995, 1989.

\bibitem{9638345}
F.~Xu, S.~A. Vorobyov, and F.~Yang, ``Transmit beamspace {DDMA} based
  automotive {MIMO} radar,'' \emph{IEEE Trans. Vehicular Technology}, vol.~71,
  no.~2, pp. 1669--1684, 2022.

\bibitem{7029030}
Y.~Liu, G.~Ye~Li, Z.~Tan, and H.~Hu, ``Noise power estimation in {SC-FDMA}
  systems,'' \emph{IEEE Wireless Communications Lett.}, vol.~4, no.~2, pp.
  217--220, 2015.

\bibitem{9903348}
D.~He, K.~Guan, B.~Ai, Z.~Zhong, J.~Kim, H.~Chung, and A.~Hrovat, ``Channel
  measurement and ray-tracing simulation for 77 {GHz} automotive radar,''
  \emph{IEEE Trans. Intelligent Transportation Systems}, vol.~24, no.~7, pp.
  7746--7756, 2023.

\bibitem{SBRA021E}
\BIBentryALTinterwordspacing
M.~A. Richards, J.~A. Scheer, and W.~A. Holm, \emph{Principles of Modern Radar:
  Basic principles}.\hskip 1em plus 0.5em minus 0.4em\relax The Institution of
  Engineering and Technology, 2010. [Online]. Available:
  \url{https://digital-library.theiet.org/doi/abs/10.1049/SBRA021E}
\BIBentrySTDinterwordspacing

\bibitem{zheng2023coarraytensor}
H.~Zheng, C.~Zhou, Z.~Shi, Y.~Gu, and Y.~D. Zhang, ``Coarray tensor
  direction-of-arrival estimation,'' \emph{IEEE Trans. Signal Process.},
  vol.~71, pp. 1128--1142, 2023.

\bibitem{7938435}
Z.~Guo, X.~Wang, and W.~Heng, ``Millimeter-wave channel estimation based on
  2-{D} beamspace {MUSIC} method,'' \emph{IEEE Trans. Wireless Communications},
  vol.~16, no.~8, pp. 5384--5394, 2017.

\bibitem{485927}
M.~Zoltowski, M.~Haardt, and C.~Mathews, ``Closed-form 2-{D} angle estimation
  with rectangular arrays in element space or beamspace via unitary {ESPRIT},''
  \emph{IEEE Trans. Signal Process.}, vol.~44, no.~2, pp. 316--328, 1996.

\bibitem{6361474}
Y.~Li, L.~Du, and H.~Liu, ``Hierarchical classification of moving vehicles
  based on empirical mode decomposition of micro-doppler signatures,''
  \emph{IEEE Trans. on Geoscience and Remote Sensing}, vol.~51, no.~5, pp.
  3001--3013, 2013.

\bibitem{10705685}
J.~Wang, J.~Wu, Y.~Qu, Q.~Xiao, Q.~Gao, and Y.~Fang, ``Multi-target device-free
  positioning based on spatial-temporal mmwave point cloud,'' \emph{IEEE Trans.
  on Mobile Computing}, vol.~24, no.~2, pp. 1163--1180, 2025.

\bibitem{8481708}
M.~Titos, A.~Bueno, L.~García, M.~C. Benítez, and J.~Ibañez, ``Detection and
  classification of continuous volcano-seismic signals with recurrent neural
  networks,'' \emph{IEEE Trans. on Geoscience and Remote Sensing}, vol.~57,
  no.~4, pp. 1936--1948, 2019.

\bibitem{10229208}
D.~Wang, Z.~Qiao, L.~Zhang, and N.~Liu, ``Total variation regularized
  self-supervised bayesian deep learning for seismic random noise
  attenuation,'' \emph{IEEE Trans. on Geoscience and Remote Sensing}, vol.~61,
  pp. 1--14, 2023.

\bibitem{9310710}
W.~Zhang, J.~Gao, Z.~Gao, and H.~Chen, ``Adjoint-driven deep-learning seismic
  full-waveform inversion,'' \emph{IEEE Trans. on Geoscience and Remote
  Sensing}, vol.~59, no.~10, pp. 8913--8932, 2021.

\bibitem{9930810}
N.~Liu, J.~Wang, J.~Gao, K.~Yu, Y.~Lou, Y.~Pu, and S.~Chang, ``{NS2NS}:
  Self-learning for seismic image denoising,'' \emph{IEEE Trans. on Geoscience
  and Remote Sensing}, vol.~60, pp. 1--11, 2022.

\bibitem{10436154}
J.~Kim, S.~Cho, S.~Hwang, W.~Lee, and Y.~Choi, ``Enhancing {LPI} radar signal
  classification through patch-based noise reduction,'' \emph{IEEE Signal
  Processing Lett.}, vol.~31, pp. 716--720, 2024.

\bibitem{7160673}
L.~Du, B.~Wang, P.~Wang, Y.~Ma, and H.~Liu, ``Noise reduction method based on
  principal component analysis with beta process for micro-{D}oppler radar
  signatures,'' \emph{IEEE J. Sel. Topics in Applied Earth Observations and
  Remote Sensing}, vol.~8, no.~8, pp. 4028--4040, 2015.

\bibitem{7891546}
N.~D. Sidiropoulos, L.~De~Lathauwer, X.~Fu, K.~Huang, E.~E. Papalexakis, and
  C.~Faloutsos, ``Tensor decomposition for signal processing and machine
  learning,'' \emph{IEEE Trans. Signal Process.}, vol.~65, no.~13, pp.
  3551--3582, 2017.

\bibitem{10284994}
Q.~Xie, X.~Pan, and F.~Zhao, ``Joint {2D-DOD and 2D-DOA} estimation in bistatic
  {MIMO} radar via tensor ring decomposition,'' \emph{IEEE Signal Process.
  Lett.}, vol.~30, pp. 1507--1511, 2023.

\bibitem{kolda2009tensor}
T.~G. Kolda and B.~W. Bader, ``Tensor decompositions and applications,''
  \emph{SIAM {R}eview}, vol.~51, no.~3, pp. 455--500, 2009.

\bibitem{4545266}
M.~Haardt, F.~Roemer, and G.~Del~Galdo, ``Higher-order {SVD}-based subspace
  estimation to improve the parameter estimation accuracy in multidimensional
  harmonic retrieval problems,'' \emph{IEEE Trans. Signal Process.}, vol.~56,
  no.~7, pp. 3198--3213, 2008.

\bibitem{10750039}
Q.~Xie, F.~Wen, X.~Xie, Z.~Wang, and X.~Pan, ``Coarray tensor train aided
  target localization for bistatic {MIMO} radar,'' \emph{IEEE Signal Process.
  Lett.}, vol.~32, pp. 46--50, 2025.

\bibitem{oseledets2011tensor}
I.~V. Oseledets, ``Tensor-train decomposition,'' \emph{SIAM J. Scientific
  Computing}, vol.~33, no.~5, pp. 2295--2317, 2011.

\bibitem{gong2020tensor}
X.~Gong, W.~Chen, J.~Chen, and B.~Ai, ``Tensor denoising using low-rank tensor
  train decomposition,'' \emph{IEEE Signal Process. Lett.}, vol.~27, pp.
  1685--1689, 2020.

\bibitem{9078764}
F.~Sedighin, A.~Cichocki, T.~Yokota, and Q.~Shi, ``Matrix and tensor completion
  in multiway delay embedded space using tensor train, with application to
  signal reconstruction,'' \emph{IEEE Signal Process. Lett.}, vol.~27, pp.
  810--814, 2020.

\bibitem{10048567}
X.~Gong, W.~Chen, L.~Sun, J.~Chen, and B.~Ai, ``An {ESPRIT}-based supervised
  channel estimation method using tensor train decomposition for mm{W}ave 3-{D}
  {MIMO-OFDM} systems,'' \emph{IEEE Trans. Signal Process.}, vol.~71, pp.
  555--570, 2023.

\bibitem{9872046}
H.~Zheng, C.~Zhou, Z.~Shi, and A.~L.~F. de~Almeida, ``{SubTTD: DOA} estimation
  via sub-{N}yquist tensor train decomposition,'' \emph{IEEE Signal Process.
  Lett.}, vol.~29, pp. 1978--1982, 2022.

\bibitem{8645667}
X.~Hu, Y.~Li, M.~Lu, Y.~Wang, and X.~Yang, ``A multi-carrier-frequency
  random-transmission chirp sequence for {TDM MIMO} automotive radar,''
  \emph{IEEE Trans. Vehicular Technology}, vol.~68, no.~4, pp. 3672--3685,
  2019.

\bibitem{zhu2022efficient}
L.~Zhu, Y.~Liu, D.~He, K.~Guan, B.~Ai, Z.~Zhong, and X.~Liao, ``An efficient
  target detection algorithm via {K}arhunen-{L}o{\`e}ve transform for frequency
  modulated continuous wave ({FMCW}) radar applications,'' \emph{IET Signal
  Process.}, vol.~16, no.~7, pp. 800--810, 2022.

\bibitem{lien2016soli}
J.~Lien, N.~Gillian, M.~E. Karagozler, P.~Amihood, C.~Schwesig, E.~Olson,
  H.~Raja, and I.~Poupyrev, ``Soli: {U}biquitous gesture sensing with
  millimeter wave radar,'' \emph{ACM Trans. Graphics}, vol.~35, no.~4, pp.
  1--19, 2016.

\bibitem{zniyed2020high}
Y.~Zniyed, R.~Boyer, A.~L. De~Almeida, and G.~Favier, ``High-order tensor
  estimation via trains of coupled third-order {CP} and {T}ucker
  decompositions,'' \emph{Linear Algebra and its Applications}, vol. 588, pp.
  304--337, 2020.

\bibitem{MDL1985}
M.~Wax and T.~Kailath, ``Detection of signals by information theoretic
  criteria,'' \emph{IEEE Trans. Acoustics, Speech, and Signal Process.},
  vol.~33, no.~2, pp. 387--392, 1985.

\bibitem{CPill}
V.~de~Silva and L.-H. Lim, ``Tensor rank and the ill-posedness of the best
  low-rank approximation problem,'' \emph{SIAM J. Matrix Analysis and
  Applications}, vol.~30, no.~3, pp. 1084--1127, 2008.

\bibitem{sorensen2013blind}
M.~S{\o}rensen and L.~De~Lathauwer, ``Blind signal separation via tensor
  decomposition with {V}andermonde factor: {C}anonical polyadic
  decomposition,'' \emph{IEEE Trans. Signal Process.}, vol.~61, no.~22, pp.
  5507--5519, 2013.

\bibitem{haardt1998simultaneous}
M.~Haardt and J.~A. Nossek, ``Simultaneous {Schur} decomposition of several
  nonsymmetric matrices to achieve automatic pairing in multidimensional
  harmonic retrieval problems,'' \emph{IEEE Trans. Signal Process.}, vol.~46,
  no.~1, pp. 161--169, 1998.

\bibitem{jiang2001almost}
T.~Jiang, N.~D. Sidiropoulos, and J.~M. Ten~Berge, ``Almost-sure
  identifiability of multidimensional harmonic retrieval,'' \emph{IEEE Trans.
  Signal Process.}, vol.~49, no.~9, pp. 1849--1859, 2001.

\end{thebibliography}
\end{document}